\documentclass[12pt,a4paper]{article}
\usepackage[T1]{fontenc}
\usepackage[utf8]{inputenc}
\usepackage{lmodern,microtype,amsmath,amssymb,amsthm,mathtools}
\usepackage[margin=25mm]{geometry}
\usepackage{setspace,booktabs,tabularx,array,enumitem,needspace}
\usepackage{graphicx,tikz,pgfplots}
\usetikzlibrary{arrows.meta,positioning}
\pgfplotsset{compat=1.18}
\usepackage[authoryear,round]{natbib}
\usepackage[dvipsnames]{xcolor}
\usepackage[colorlinks=true,linkcolor=MidnightBlue,citecolor=MidnightBlue,urlcolor=MidnightBlue]{hyperref}
\usepackage[font=small,labelfont=bf]{caption}
\usepackage[section]{placeins}
\hypersetup{pdftitle={Costly Comparison and Informational Inertia},pdfauthor={Georgy Lukyanov and Arina Azova}}
\setlist{nosep,leftmargin=*}
\newtheorem{proposition}{Proposition}
\newcommand{\E}{\mathbb E}
\newcommand{\Pp}{\mathbb P}

\newcommand{\PP}{\mathcal P}
\newcommand{\1}{\mathbf 1}

\newcommand{\mustar}{\mu^*}
\newcolumntype{Y}{>{\raggedright\arraybackslash}X}
\newcolumntype{L}[1]{>{\raggedright\arraybackslash}p{#1}}
\definecolor{firmblue}{RGB}{35,75,111}
\definecolor{learnred}{RGB}{163,71,43}
\definecolor{lightblue}{RGB}{237,243,248}

\newcommand{\HiddenInformationCoordinates}{(0.0001,5.0000000041666667e-9)(0.00010714910692583925,5.7404655629946243e-9)(0.00011480931115004934,6.5905889706137319e-9)(0.00012301715156498585,7.5666097991236145e-9)(0.00013181177926748843,8.687172589408374e-9)(0.00014123514430817239,9.9736830104541992e-9)(0.00015133219579162701,1.1450716763410818e-8)(0.00016215109628199084,1.314648904153076e-8)(0.00017374345153661092,1.5093393513895755e-8)(0.00018616455666360694,1.7328621128925511e-8)(0.00019947365987750281,1.9894870558430482e-8)(0.0002137342451110304,2.2841163853544139e-8)(0.00022901433483115332,2.6223782893692267e-8)(0.000245386814503732,3.0107344517219855e-8)(0.00026292978025451463,3.4566034871478803e-8)(0.00028172691138478407,3.9685026561688542e-8)(0.00030186786951854668,4.5562105669817407e-8)(0.00032344872628518051,5.2309539723800799e-8)(0.0003465724215775732,6.0056222300196533e-8)(0.0003713492545716243,6.8950135227804038e-8)(0.00039789740984925673,7.9161175426788975e-8)(0.00042634352113452495,9.088440038335493e-8)(0.00045682327533182022,1.0434375425704918e-7)(0.0004894820597474131,1.1979634579914578e-7)(0.00052447565558155605,1.3753735980160571e-7)(0.0005619709809990779,1.5790569589823443e-7)(0.00060214688732288978,1.8129044243402651e-7)(0.00064519501214821598,2.0813830907072752e-7)(0.00069132069344687351,2.3896216011108528e-7)(0.0007407439490218439,2.7435081155096589e-7)(0.00079370052598409974,3.1498027900914748e-7)(0.0008504430252576516,3.6162669140033704e-7)(0.00091124210647666325,4.151811170372395e-7)(0.00097638777902194989,4.7666658538021303e-7)(0.0010461907853550562,5.472576295960659e-7)(0.0011209840832483666,6.2830272324214106e-7)(0.0012011244339814312,7.2135003967791798e-7)(0.0012869941040791453,8.2817702628010223e-7)(0.0013790026887090103,9.5082435841116177e-7)(0.0014775890654350157,1.091634921757688e-6)(0.0015832234876474739,1.2532985677121878e-6)(0.0016964098276543933,1.4389034967550664e-6)(0.0018176879801338512,1.6519952514095667e-6)(0.001947636437411748,1.896644445710284e-6)(0.0020868750488489202,2.1775245250228771e-6)(0.0022360679774997897,2.5000010416674479e-6)(0.0023959268681457009,2.8702341517917831e-6)(0.0025672142418143487,3.2952962915166201e-6)(0.00275074713297703,3.7833072803593707e-6)(0.0029474009867730156,4.3435894328763447e-6)(0.0031581138348506598,4.9868456417084569e-6)(0.003383890769743856,5.7253638340762958e-6)(0.0036258087391264492,6.5732517076342899e-6)(0.0038850216828130231,7.5467062300904331e-6)(0.0041627660370093657,8.6643230511847056e-6)(0.0044603666320676865,9.9474517381919639e-6)(0.0047792430118786606,1.1420603621651744e-5)(0.0051209162050435665,1.3111920043267581e-5)(0.0054870159801247608,1.5053709952083846e-5)(0.0058792886195817667,1.7283067120211337e-5)(0.0062996052494743658,1.9842578770881137e-5)(0.0067499707646650709,2.2781139158567792e-5)(0.0072325333920938664,2.6154883646766767e-5)(0.0077495949377416857,3.0028261132129793e-5)(0.0083036217661602648,3.4475265308150627e-5)(0.0088972565649403241,3.9580848298101017e-5)(0.0095333309502341605,4.5442543673620866e-5)(0.010214878973460528,5.2172329879101555e-5)(0.010945151593618292,5.9898769679550098e-5)(0.011727632184241263,6.8769466529206703e-5)(0.012566053148961809,7.8953884821186883e-5)(0.013464413724938879,9.0646587941340836e-5)(0.014426999059072136,0.00010407095604628742)(0.015458400647995026,0.00011948345467014647)(0.016563538239344818,0.00013717853581389224)(0.017747683298777852,0.0001574942652890888)(0.019016484154666796,0.00018081878400827683)(0.020375992940419204,0.00020759772690700377)(0.021832694462931231,0.00023834274155770743)(0.023393537134877971,0.00027364126965069553)(0.02506596611838631,0.00031416777878397202)(0.026857958838184386,0.0003606966598961612)(0.028778063033624082,0.00041411703774254722)(0.030835437531083288,0.00047544977868488575)(0.0330398959312308,0.00054586702247063422)(0.035401953419540503,0.00062671461345981275)(0.037932876923339259,0.00071953786289350462)(0.040644738854635786,0.00082611113841714537)(0.043550474695081832,0.00094847185148487723)(0.046663944697743799,0.0010889594990005047)(0.05,0.0012502605143665344)}
\newcommand{\ObservedInformationCoordinates}{(0.0001,2.7466557279206128e-5)(0.00010714910692583925,2.9430266590414625e-5)(0.00011480931115004934,3.1534377762570555e-5)(0.00012301715156498585,3.3788930375368207e-5)(0.00013181177926748843,3.6204682059906175e-5)(0.00014123514430817239,3.8793159879034265e-5)(0.00015133219579162701,4.1566715387763367e-5)(0.00016215109628199084,4.453858363784153e-5)(0.00017374345153661092,4.772294640963035e-5)(0.00018616455666360694,5.1134999974829573e-5)(0.00019947365987750281,5.4791027715496411e-5)(0.0002137342451110304,5.8708477948299628e-5)(0.00022901433483115332,6.2906047328155212e-5)(0.000245386814503732,6.7403770232437945e-5)(0.00026292978025451463,7.2223114555988764e-5)(0.00028172691138478407,7.7387084378289427e-5)(0.00030186786951854668,8.2920329997613297e-5)(0.00032344872628518051,8.8849265862856058e-5)(0.0003465724215775732,9.5202196972288716e-5)(0.0003713492545716243,0.00010200945434985758)(0.00039789740984925673,0.00010930354025409885)(0.00042634352113452495,0.00011711928382247259)(0.00045682327533182022,0.00012549400790520547)(0.0004894820597474131,0.00013446770789783735)(0.00052447565558155605,0.00014408324344088914)(0.0005619709809990779,0.00015438654391873024)(0.00060214688732288978,0.00016542682875816988)(0.00064519501214821598,0.00017725684360090563)(0.00069132069344687351,0.0001899331135031428)(0.0007407439490218439,0.00020351621440089567)(0.00079370052598409974,0.00021807106417117819)(0.0008504430252576516,0.0002336672347180136)(0.00091124210647666325,0.00025037928661851204)(0.00097638777902194989,0.00026828712797880661)(0.0010461907853550562,0.00028747639927308175)(0.0011209840832483666,0.00030803888607201679)(0.0012011244339814312,0.00033007296171051449)(0.0012869941040791453,0.00035368406209947865)(0.0013790026887090103,0.00037898519505361954)(0.0014775890654350157,0.00040609748668787108)(0.0015832234876474739,0.0004351507676301698)(0.0016964098276543933,0.00046628420200936494)(0.0018176879801338512,0.00049964696240531877)(0.001947636437411748,0.00053539895419538238)(0.0020868750488489202,0.00057371159299912068)(0.0022360679774997897,0.00061476863921331924)(0.0023959268681457009,0.00065876709394404682)(0.0025672142418143487,0.0007059181609842105)(0.00275074713297703,0.00075644827985622171)(0.0029474009867730156,0.0008106002353429782)(0.0031581138348506598,0.00086863434936956966)(0.003383890769743856,0.0009308297615765152)(0.0036258087391264492,0.00099748580544693567)(0.0038850216828130231,0.0010689234874193224)(0.0041627660370093657,0.0011454870770394923)(0.0044603666320676865,0.0012275458168855286)(0.0047792430118786606,0.0013154957617443096)(0.0051209162050435665,0.0014097617573347103)(0.0054870159801247608,0.0015107995697687351)(0.0058792886195817667,0.0016190981779267212)(0.0062996052494743658,0.0017351822420065559)(0.0067499707646650709,0.0018596147627011411)(0.0072325333920938664,0.0019929999467762286)(0.0077495949377416857,0.0021359862962771549)(0.0083036217661602648,0.0022892699402048871)(0.0088972565649403241,0.0024535982292885138)(0.0095333309502341605,0.0026297736164649764)(0.010214878973460528,0.002818657847882738)(0.010945151593618292,0.003021176491703229)(0.011727632184241263,0.0032383238347155973)(0.012566053148961809,0.0034711681798448416)(0.013464413724938879,0.0037208575810650007)(0.014426999059072136,0.0039886260560787101)(0.015458400647995026,0.0042758003214511656)(0.016563538239344818,0.0045838070997587818)(0.017747683298777852,0.0049141810538101256)(0.019016484154666796,0.0052685734092116158)(0.020375992940419204,0.0056487613335910087)(0.021832694462931231,0.0060566581487841305)(0.023393537134877971,0.0064943244613826657)(0.02506596611838631,0.0069639803074068982)(0.026857958838184386,0.0074680184187118243)(0.028778063033624082,0.0080090187322996136)(0.030835437531083288,0.0085897642792819016)(0.0330398959312308,0.0092132586081511045)(0.035401953419540503,0.0098827449176844934)(0.037932876923339259,0.010601727098699781)(0.040644738854635786,0.011373992911582835)(0.043550474695081832,0.012203639558708758)(0.046663944697743799,0.013095101948410184)(0.05,0.014053183991018092)}

\title{\vspace{-12mm}Costly Comparison and Informational Inertia\thanks{We are grateful to Alexis Belianin, Markus Gebauer, Vitalijs Jascisens, Ella Khromova, Yukio Koriyama, Jean-Baptiste Michau, Fabian Slonimczyk, and Anna Yurko for helpful comments and discussions. We also thank participants at the ICEF research seminar. All remaining errors are our own. Financial support from the French National Research Agency (ANR) under grant ANR-17-EURE-0010 (Investissements d'Avenir program) is gratefully acknowledged.}}
\author{Georgy Lukyanov\thanks{Toulouse School of Economics, University of Toulouse Capitole, France. Corresponding author: \href{mailto:georgy.lukyanov@tse-fr.eu}{georgy.lukyanov@tse-fr.eu}.}
\and Arina Azova\thanks{International College of Economics and Finance, HSE University, Russia. Email: \href{mailto:asazova@edu.hse.ru}{asazova@edu.hse.ru}.}}
\date{October 2026}
\begin{document}
\maketitle
\begin{abstract}
\singlespacing
A price cut can attract consumers while changing what their purchases reveal to later buyers. We study this trade-off in a duopoly with common uncertainty about quality. Consumers have a default seller and may pay to obtain access to the rival together with a private signal. Firms observe purchases and retain their prices between random opportunities to revise them. In a specified pricing experiment, a cut raises current revenue and remains profitable over two arrivals, but loses over three for sufficiently patient firms. The same cut remains profitable when quality is independently redrawn for each buyer. Using analytical arguments and rigorous numerical bounds, we then construct stationary Markov equilibria on three finite price menus in which an informative purchase leads with positive probability to uninformative purchases and unequal prices. These equilibria require infrequent repricing. Finally, at fixed prices, we introduce a small share of consumers who compare cheaply. Expected exit time from a fixed belief interval is of order the inverse square of this share when defaults are hidden, and of order its inverse when defaults are observed.
\end{abstract}
\vspace{2mm}
\clearpage

\section{Introduction}\label{sec:intro}

A firm considering a price cut usually weighs the gain in sales against
the lower revenue from each customer. When consumers learn from previous
purchases, there is another consideration. The cut may induce buyers to
compare products, making their decisions more informative to those who
arrive later. The firm will then face a different demand curve while the
lower price is still in place. The current increase in sales need not be
a reliable guide to the profitability of the cut.

This paper studies that incentive when comparison is costly and prices
cannot be revised after every purchase. Consumers begin with a default
seller. To consider the rival, they must incur a cost which also provides
a private signal about the products' relative quality. They subsequently
decide whether to stay with the default, switch, or make no purchase.
Their final decisions are publicly observed, but the comparison activity,
private signal, and default are not. A purchase can therefore convey little
information even when the two firms continue to sell. The relevant question
for a firm is how changing its price alters both demand and the information
on which future demand will depend.

We focus on a common quality ranking which is initially unknown to firms
as well as consumers. Prices do not signal information privately held by
the sellers. Instead, they affect the incentives of buyers to obtain and
act upon information. A lower price can make it worthwhile for some
consumers attached to the rival to compare the products. This creates an
immediate sales gain. It can also create an unfavorable continuation: if
the comparison results in a purchase from the rival, subsequent consumers
may find comparison unattractive, leaving the deviating firm with its
default customers at the reduced price. The corresponding gain after a
favorable purchase need not compensate for this loss.

Our first result makes this mechanism explicit in a short pricing
experiment. Starting from a particular public belief, we hold one firm's
price fixed and compare the other firm's current price with a specified
cut. The cut increases revenue at the first arrival and is profitable over
two arrivals for every discount factor. Over three arrivals, however, it
becomes unprofitable when the firm is sufficiently patient. The reason is
the asymmetric response of subsequent demand to the purchases generated
by the cut. There is no adverse drift in expected quality: public beliefs
remain a martingale. It is the firm's revenue at the installed price which
responds unfavorably to the distribution of the new information.

To identify the role of learning, we compare this experiment with one in
which quality is independently redrawn for each arriving consumer. The
current buyer faces exactly the same choice problem in the two economies.
Default shares, comparison and switching costs, signal accuracy, values,
and the two price choices are unchanged. What changes is whether the
current purchase is informative about the quality relevant to the next
buyer. In the independent-quality economy the cut's initial advantage
persists at every horizon. This comparison isolates an effect of persistent
quality which a comparison with fully known quality would not isolate:
revealing quality would also change the current buyer's information problem.

We next ask whether the pricing incentive can be embedded in equilibrium.
Firms receive independent Poisson opportunities to reset prices chosen
from a finite menu. We construct stationary Markov equilibria with explicit
reset targets and a positive-probability route into an informationally
silent region. From a symmetric prior and specified inherited prices, a
purchase from one firm before either can reprice changes the public belief
and brings the state into that region. Subsequent buyers purchase their
defaults without comparing. Each firm eventually installs its prescribed
price, and the resulting price gap is strictly positive. The construction
works on a cent menu and on two finer menus, with appropriately adjusted
target prices.

For the cent-menu targets, the role of learning also survives the full
dynamic comparison. If quality is independently redrawn for each buyer,
the same cut is profitable even when all revenue after the next reset is
ignored. The proposed target pair therefore cannot remain absorbing in
that control economy. The equilibrium construction establishes this
pricing incentive under infrequent repricing and along a specified
reachable path. It does not rank equilibrium price dispersion: a separate
control with costly access but no private signal also supports unequal
prices.

Exact silence raises a further question. If a small fraction of consumers
can compare cheaply, purchases may remain informative. Our third result
therefore studies the duration of informational inertia, holding prices
fixed. We place cheap comparison among consumers with low switching costs
and ask how long the public posterior takes to leave a fixed interval
around the silent belief. When defaults are hidden, expected exit time is
of order the inverse square of the cheap-inspector share. When the default
is observed together with the purchase, it is of order the inverse of that
share. An informed purchase is difficult to distinguish from ordinary
default purchases in the first case. In the second, a switch can be
identified and provides a substantial piece of information. This is a
comparison of learning under specified prices; the firms' equilibrium
response to the perturbed consumer population is a separate problem.

The interaction between pricing and learning is central to
\citet{Bose2006,Bose2008}, who study a monopolist whose price determines
the information conveyed by a buyer's decision. In their binary model, a
low pooling price suppresses learning, whereas a higher separating price
reveals the signal. The seller can choose a new price before the next buyer
and benefits from the resulting public information when future prices
are chosen optimally. Here a cut can activate costly comparison, and its
effect on future demand is evaluated while the chosen price remains
installed. Costly comparison determines which buyers transmit information, while
price persistence determines the firm's exposure to the resulting demand.

\citet{PapanastasiouSavva2017} also analyze the interaction between social
learning and price commitment. Their consumers can delay buying in order
to learn from early buyers' reviews, and the firm chooses between a
preannounced price path and responsive pricing. Our consumers make a
single decision upon arrival, and their purchases themselves generate
the public evidence. Costly rival access affects whether a consumer's private signal enters
that evidence, and the installed price determines the revenue earned as
subsequent consumers learn.

On the duopoly side, \citet{ArieliKorenSmorodinsky2022} examine strategic
pricing when both firms and consumers are uncertain about quality and
observe past prices and purchases. Their analysis includes farsighted firms who can choose prices before
each consumer. We instead examine costly acquisition and switching while
a price remains installed. Our conclusions concern a specified pricing
incentive, a reachable equilibrium, and the duration of information spells.

Our treatment of consumer defaults is related to the search environment
of \citet{GarciaShelegia2018}. In their model a consumer observes a
predecessor's purchase and uses it to decide where to begin searching.
First visits therefore respond to past market outcomes. Here the default
is an exogenous, privately known consideration advantage, and the purchase
history affects beliefs about a common quality ranking. This distinction
allows us to hold the distribution of defaults fixed when changing the
information in purchases. Heterogeneous actions can persist without learning, as the type-specific
herds and confounding in \citet{SmithSorensen2000} demonstrate. In our
silent region, neither default group responds to a private signal;
informational responses do not cancel across types.

Cost heterogeneity also matters for endogenous information acquisition.
\citet{MuellerFrankPai2016} show, in a sequential-search model, that
asymptotic learning occurs exactly when search costs are not bounded
away from zero. Their agents sample actions to learn their qualities and
choose among the sampled actions. Here comparison provides rival access
and one noisy signal. Our rare-inspection result holds the cheap cost
strictly positive and varies the share of cheap types. It concerns exit
from a fixed belief interval under specified prices, rather than eventual
revelation of quality.

Recent work also emphasizes that the observation technology matters for
pricing. \citet{ChenChenIshida2025} study anonymous ratings and the
incentive to attract favorable evaluations through prices, while
\citet{XuYang2026} study pricing and recommendations when consumer
feedback informs the seller. The observation in our model is a purchase,
with or without the buyer's default. Our duration result makes the
difference between these two observations explicit within the same
consumer decision problem.

The analysis uses a binary signal and three specified finite price menus.
These restrictions permit verification of all reset deviations in the
equilibrium construction. The broader cost checks and duration comparison
concern fixed-price experiments; their assumptions and implications are
stated separately from the equilibrium result.

Section~\ref{sec:model} presents the environment.
Section~\ref{sec:incentive} studies the price-cut experiment, and
Section~\ref{sec:equilibrium} constructs the reachable equilibrium.
Section~\ref{sec:duration} introduces rare inspection and compares the
two observation regimes. Section~\ref{sec:conclusion} concludes.

\section{Environment}\label{sec:model}

\subsection{Quality, consumers, and comparison}

Two firms, $A$ and $B$, sell products with a common but unknown quality
ranking. The state $\theta\in\{A,B\}$ is drawn once and remains fixed.
The product values are $(v_A,v_B)=(1.5,0.5)$ in state $A$ and
$(0.5,1.5)$ in state $B$. Let $\mu$ denote the public probability of
state $A$. The expected product values are then
$R_A(\mu)=0.5+\mu$ and $R_B(\mu)=1.5-\mu$. Firms have zero production
cost and no private information about $\theta$.

Consumers arrive according to a Poisson process of rate $\lambda$.
Each consumer demands at most one unit and makes her decision at arrival;
there is no option to return or wait for a later price. She privately knows
her default $J\in\{A,B\}$, her switching cost $K$, and her comparison
cost $C$. A default represents a product that is already available for
consideration, for example through familiarity. It need not be the
supplier in a publicly observed previous transaction. Defaults are equally
likely and independent across buyers and of quality. The distribution of
private types is common knowledge.

Both posted prices are known. Without comparing, a consumer can buy her
default or take an outside option of zero. Paying $C$ gives access to the
rival and a binary signal $s\in\{+,-\}$, with
$\Pp(+\mid A)=\Pp(-\mid B)=q=0.75$. After comparing she can purchase
either product or exit. Signals are independent across consumers and of
private types conditional on quality. The fee buys access and information together. It may represent the effort
needed to evaluate and activate an unfamiliar alternative; the separate
cost $K$ is incurred only if that alternative is used. Prices are already
known, and the rival cannot be purchased without paying $C$.

Let $e\in\{0,1\}$ indicate comparison and let $y\in\{A,B,0\}$ be
the final choice, where $0$ denotes no purchase. The realized payoff is
\begin{equation}
 U(y,e)=
 \begin{cases}
 v_y(\theta)-p_y-K\1\{y\ne J\}-eC,&y\in\{A,B\},\\
 -eC,&y=0.
 \end{cases}
 \label{eq:utility}
\end{equation}
If $e=0$, the feasible choices are $J$ and $0$. If $e=1$, all three
choices are feasible. In particular, a consumer who exits after comparing
still pays $C$. Switching costs are incurred only when the rival is
purchased. Consumption payoffs and reviews are not subsequently added to
the public history.

Write $\pi_s(\mu)$ for the probability of signal $s$ and $\mu_s$ for
the associated posterior. For a consumer with default $j$ and rival $k$,
the gross gain from comparing is
\begin{equation}
 B_j(K;\mu,p)=
 \sum_s\pi_s(\mu)\max\{R_j(\mu_s)-p_j,
                 R_k(\mu_s)-p_k-K,0\}
 -\max\{R_j(\mu)-p_j,0\}.
 \label{eq:inspection}
\end{equation}
She compares exactly when $B_j>C$, and then chooses the product or
outside option with the highest expected payoff. We favor the default
in a switching tie and purchase in a zero-surplus tie. These conventions
define decisions at every price, including deviations; the equilibrium
targets below have strictly positive default purchase surplus.

Table~\ref{tab:primitives} gives the central parameterization. Later
sections change the comparison-cost law explicitly. Unless stated
otherwise, the switching-cost marginal is exponential with mean $0.14$
within each default group. Types are independent of quality and are
independently drawn across arrivals. When comparison costs are
heterogeneous, we allow them to be correlated with switching costs.

\begin{table}[tb]
\centering\small\singlespacing
\caption{Central primitives}\label{tab:primitives}
\begin{tabularx}{\textwidth}{@{}L{.35\textwidth}Y@{}}
\toprule
Object & Specification\\\midrule
Product values & $(1.5,0.5)$ in state $A$; $(0.5,1.5)$ in state $B$\\
Default shares & One half for each firm; independent of quality\\
Signal accuracy & $q=0.75$\\
Comparison cost & $C=0.3$; rival access and a signal\\
Switching cost & Exponential, mean $0.14$; lower endpoint zero\\
Outside option & Zero; available before and after comparison\\
Arrival and discount rates & $\lambda=1$, $\rho=0.25$\\
Public observations & Arrival times, purchase outcomes, posted prices, reset events\\
Private information & Default, costs, comparison decision, and signal\\
\bottomrule
\end{tabularx}
\caption*{The equilibrium menus and reset-rate restriction are specified in Section~\ref{sec:equilibrium}. The fixed-price experiments impose no equilibrium restriction on reset rates.}
\end{table}

\subsection{Public information}

Let $\tau_n$ be the time of the $n$th consumer arrival and
$Y_n\in\{A,B,0\}$ its outcome. In the baseline, the public observes
$Y_n$ but not $J_n$. The observed-default experiment instead reveals
$O_n=(J_n,Y_n)$; otherwise $O_n=Y_n$. For a state
$x=(\mu,p_A,p_B)$, let $\ell_\theta(o\mid x)$ be the probability of
outcome $o$ under quality state $\theta$, integrating optimal consumer
choices over types and signals. Bayes' rule gives
\begin{equation}
 T_o(x)=\frac{\mu\ell_A(o\mid x)}
 {\mu\ell_A(o\mid x)+(1-\mu)\ell_B(o\mid x)}.
 \label{eq:bayes}
\end{equation}
This expression is used only for outcomes with positive predictive
probability. A zero-probability outcome does not generate a reachable
successor. Appendix~\ref{app:consumer} derives the likelihoods from
the consumer problem, keeping no purchase as a possible outcome.

We call a state \emph{informationally silent} if the distributions of
the public observation coincide under $\theta=A$ and $\theta=B$.
An arrival then leaves the public belief unchanged. Silence does not
require all consumers to choose the same product. In the region used for
the equilibrium construction, every consumer buys her default without
comparing, so each firm sells with probability one half in either state.

At the central comparison cost, every consumer who compares purchases
after either signal. Appendix~\ref{app:consumer} shows that
$C>2-\sqrt3$ is sufficient for this property. Non-purchases therefore
come from non-comparing consumers and carry no information about quality.
This simplifies the calculations, but we do not impose it by deleting
the outside option. Deviations can cause some consumers not to purchase,
and their lost demand still matters for profits. For the cheap types in
Section~\ref{sec:duration}, participation is checked directly on the
specified belief interval.

\subsection{Installed prices and equilibrium}

Each firm receives independent Poisson opportunities of rate $\alpha$
to reset its price. These clocks and consumer arrivals are independent of
quality and private types. Let $T_m$ denote reset-event times. We write
$p_i(t-)$ for firm $i$'s price just before an event and $p_i(t)$ for
its price immediately afterwards. At an arrival prices do not change;
at a reset only the resetting firm chooses a new price. Simultaneous
events have probability zero. This opportunity structure follows the
familiar Poisson formulation of price adjustment \citep{Calvo1983}.
Its role here is to make a chosen price persist while new purchases
change public information. Setting $\lambda=1$ fixes the time unit;
the relevant timing parameters are $\rho/\lambda$ and $\alpha/\lambda$.

The feasible menu $\PP$ is finite and is the same at every belief.
A stationary Markov strategy specifies a distribution over reset prices
as a function of $(\mu,p_A,p_B)$. Prices, reset events, and strategies
are public. A reset does not directly reveal quality information because
the firm possesses no private quality signal. Its choice can nevertheless
alter the information conveyed by subsequent consumer decisions.

For any strategy profile, normalize firm $i$'s discounted profit as
\begin{equation}
 V_i(x)=\frac{\rho}{\lambda}
 \E_x\!\left[\sum_{n\ge1}e^{-\rho\tau_n}
 p_i(\tau_n-)\1\{Y_n=i\}\right].
 \label{eq:value}
\end{equation}
A perpetual expected revenue of $u$ per arrival then has value $u$.
A stationary Markov equilibrium consists of sequentially optimal consumer
choices, Bayesian updating, and firms' stationary strategies which are
optimal at every reachable reset state, allowing deviations to arbitrary
feasible continuation strategies. The state space is closed under all
feasible price choices and positive-probability observations from the
specified initial state. It is countable, not generally finite, because
the set of posterior beliefs can be infinite.

We will construct equilibrium by specifying reset choices on a closed
region and proving that they are optimal against any admissible
continuation after leaving that region. A discounted-game argument then
supplies equilibrium behavior elsewhere. Before undertaking that
verification, it is useful to see the particular pricing incentive which
motivates it.

\section{A price cut and its future demand}\label{sec:incentive}

\subsection{The comparison}

Consider belief $\mustar=0.414326273258635\ldots$ and prices
$(p_A,p_B)=(0.91,1.08)$. Section~\ref{sec:equilibrium} derives this
belief from a purchase at an interior prior. At the stated prices every
consumer purchases her default without comparing. Expected revenue per
arrival is therefore $0.455$ for $A$ and $b=0.54$ for $B$.
Although $B$ is perceived as the better product, the quality ranking is
far from known.

Hold firm $A$'s price at $0.91$ and let $B$ cut its price to $0.85$.
The cut makes comparison worthwhile for some consumers, and its
probability of a sale rises to approximately $0.660845$. Its current
expected revenue is $0.561718$, which exceeds $0.54$. A firm that
evaluated the cut solely at the current public belief would therefore
prefer it. To see whether this preference survives, keep the price pair
fixed through the next arrivals and update public beliefs after purchases.

The first purchase is either $A$ or $B$. A purchase from $B$ lowers
$\mu$, the probability that $A$ is better, from about $0.414326$ to
$0.359925$. Firm $B$'s next-arrival revenue then rises to about
$0.571659$. A purchase from $A$ instead raises $\mu$ to about
$0.520327$, where further comparison becomes unattractive at the
installed price pair. Each firm
then retains its default customers. Firm $B$ earns only $0.425$ per
arrival, because it has cut the price charged to the half of consumers
who would have bought from it without the cut.

Figure~\ref{fig:incentive} shows these branches and the first three
expected revenues. The favorable branch does not offset the unfavorable
one: revenue at the second arrival is below the stay revenue. Further
observations matter as well. In particular, after an initial $B$ purchase,
some consumers may decline to buy at the next arrival. Such an outcome
does not change the posterior, but it must be counted when comparing
revenues across arrivals. Appendix~\ref{app:incentive} reports the full
tree, including this branch.

\begin{figure}[tb]
\centering\singlespacing
\begin{tikzpicture}[font=\small,>=Stealth]
\node[draw=firmblue,fill=lightblue,rounded corners,align=center,text width=4.4cm] (root) at (0,0)
{$\mu^*=0.414326$\\Cut to $p_B=0.85$\\Revenue: $0.561718$};
\node[draw=firmblue,align=center,text width=5.1cm] (left) at (-3.2,-2.8)
{$A$ purchase\\Probability: $0.339155$\\$\mu'=0.520327$\\Next revenue: $0.425000$\\\textbf{Silent}};
\node[draw=firmblue,align=center,text width=5.1cm] (right) at (3.2,-2.8)
{$B$ purchase\\Probability: $0.660845$\\$\mu'=0.359925$\\Next revenue: $0.571659$\\\textbf{Informative}};
\draw[->,thick,firmblue] (root.south west)--(left.north);
\draw[->,thick,firmblue] (root.south east)--(right.north);
\end{tikzpicture}
\vspace{4mm}

\begin{tikzpicture}
\begin{axis}[width=.85\textwidth,height=5.2cm,xmin=.85,xmax=3.15,
 ymin=.49,ymax=.58,xtick={1,2,3},ytick={.50,.52,.54,.56,.58},
 xlabel={Buyer arrival},ylabel={Expected revenue of $B$},
 tick label style={font=\small},label style={font=\small},
 legend style={font=\footnotesize,at={(.5,1.03)},anchor=south,legend columns=1,draw=none},
 grid=major,grid style={gray!18}]
\addplot[learnred,thick,mark=*] coordinates {(1,.5617180890479141)(2,.5219188793879708)(3,.5024843674056841)};
\addlegendentry{Cut, persistent quality}
\addplot[black,dashed,thick] coordinates {(1,.54)(3,.54)};
\addlegendentry{Stay at $1.08$}
\addplot[firmblue,dashdotted,thick,mark=square*] coordinates {(1,.5617180890479141)(2,.5617180890479141)(3,.5617180890479141)};
\addlegendentry{Cut, independent quality}
\end{axis}
\end{tikzpicture}
\caption{The cut's current gain and its subsequent demand. The rival price is
$0.91$, comparison costs are $0.3$, and prices remain fixed. In the upper
panel, $\mu'$ always denotes the probability that $A$ is better. The lower
panel averages over all histories, including no purchase. Lines join the
three arrival values for readability; they do not extrapolate beyond them.}
\label{fig:incentive}
\end{figure}

Let $a_n$ be firm $B$'s expected revenue at arrival $n+1$ following
the cut. With an interarrival discount factor $\delta$, the discounted
gain over $N$ arrivals, apart from the common discount to the first
arrival, is
\begin{equation}
 G_N(\delta)=\sum_{n=0}^{N-1}\delta^n(a_n-b),
 \qquad \delta=\frac{\lambda}{\lambda+\rho}.
 \label{eq:gain}
\end{equation}
The comparison has no terminal payoff. An identical affine terminal
value can be added without changing it, as shown in the appendix.

\Needspace{14\baselineskip}
\begin{proposition}\label{prop:incentive}
At the central primitives, start at $\mustar$, hold $p_A=0.91$, and
compare staying at $p_B=1.08$ with cutting to $0.85$. Keep the chosen
prices through the relevant arrivals, with no terminal payoff. The cut
raises current revenue and is profitable over two arrivals for every
$\delta\in[0,1]$. Over three arrivals it is unprofitable exactly when
$\delta>\delta^*=0.557128884106174\ldots$.

If quality is instead independently drawn for each arrival with probability
$\mustar$ of state $A$, keeping the current consumer problem and the price
choices unchanged, the cut yields $a_0>b$ at every arrival and is profitable
at every finite horizon.
\end{proposition}

The first three expected revenues after the cut are
$a_0=0.5617180890\ldots$, $a_1=0.5219188793\ldots$, and
$a_2=0.5024843674\ldots$. Write $g_n=a_n-b$. The signs are
$g_0>0$ and $g_1,g_2<0$, but
$g_0+g_1>0$. Thus $G_2$ remains positive even without discounting
the second arrival. By contrast, $G_3$ is strictly decreasing on
$[0,1]$, positive at zero, and negative at one. Its unique root gives
the stated threshold. At the central discount rate, $\delta=0.8$,
and the cut's weighted average revenue is $0.533132454$, below $0.54$.
Equivalently, the three-arrival reversal occurs when
$\rho/\lambda<0.7949168110\ldots$.

\subsection{What persistent quality changes}

The independent-quality control holds the first consumer's incentives
fixed. She faces the same public probability of the quality states,
the same private signal, the same costs, and the same posted prices.
Her purchase is consequently distributed in exactly the same way.
However, it conveys no information about the fresh quality draw facing
the next consumer. Each arrival therefore reproduces the initial sales
gain. In the persistent-quality economy, the purchase changes future
beliefs and hence the next consumer's incentive to compare or purchase.
This is the entire difference between the two experiments.

The adverse continuation arises because the cut changes which consumers
compare and how their choices inform later buyers. Public beliefs are a
martingale, so the firm does not expect the quality of its product to be
reassessed downward on average. Revenue at a fixed price is a nonlinear function
of that belief, partly because the decision to compare changes with it.
The particular information generated by the cut can therefore reduce
expected future revenue. This does not imply that revenue falls after
every history. In fact, the calculation has histories at which the
expected next-arrival change is positive.

The freedom to reprice matters for the same reason. After a price spell
ends, information may enable a firm to select a more attractive price.
Proposition~\ref{prop:incentive} evaluates returns while the lower price
remains installed. It cannot be used on its own to dispose of a nonlinear
continuation value in the dynamic game. The equilibrium verification in
the next section allows an upper bound on every feasible continuation
after the first reset, rather than assuming that learning has no value
at that date.

\subsection{Cost and duration checks}

The sign reversal is not confined to a single value of the comparison
cost. Holding $\mustar$ fixed, it survives for every homogeneous
$c\in[0.295,0.305]$ and every $\delta\in[0.8,1]$: the stay price is
silent, the cut raises current revenue, and its three-arrival value is
lower. A separate calculation permits arbitrary heterogeneity in
$C\in[0.299,0.301]$ at $\delta=0.8$, including correlation with $K$,
while preserving the switching-cost marginal and independence of types
from quality. These are distinct claims. The wider range of homogeneous
economies does not imply robustness to an arbitrary cost distribution
on that entire range.

Appendix~\ref{app:incentive} also considers a price spell ending at an
independent exponential time. For the central law, the cut loses over a
spell with mean length $1.3333$ arrival-time units, but pays when the mean
length is $0.5714$. Thus the within-spell incentive can operate over a
moderate number of arrivals. This calculation concerns revenues before
the spell ends, or an identical affine terminal value. It is not an
equilibrium result for firms resetting at either of those rates.

The next step is consequently more demanding than this short experiment.
To support a price as an equilibrium reset choice, we must check every
price in the menu and allow the market to continue after the next reset.
We must also explain how the relevant public belief can be reached.

\section{A reachable pricing equilibrium}\label{sec:equilibrium}

\subsection{Entry into a silent region}

Let $\PP_d=\{0,d,2d,\ldots,1.5\}\cup\{2\}$, where the menu tick
is one of $d=0.01$, $0.005$, or $0.001$. The price $2$ makes no sales
in the central economy. It ensures that prices above buyers' values
remain feasible; the proposed targets are not the largest available
prices. Begin with the symmetric prior $\mu_0=1/2$ and inherited
prices $(0.79,0.99)$. These inherited prices are initial conditions.
We do not model the decisions that installed them before the game begins.

At this initial state, consumers attached to $A$ do not compare.
Those attached to $B$ compare when $K<0.1$. After comparing, they
buy $A$ following $+$ and $B$ following $-$. Let
$H=\Pp(K<0.1)=1-e^{-5/7}$. The quality-state likelihoods of a
$B$ purchase are $1/2-3H/8$ under $A$ and $1/2-H/8$ under $B$.
A first purchase from $B$ therefore gives the posterior
\begin{equation}
 \mustar=\frac{1/4-3H/16}{1/2-H/4}
        =0.414326273258635\ldots,
 \qquad H=1-e^{-5/7}.
 \label{eq:entry}
\end{equation}
This is the belief used in Section~\ref{sec:incentive}. The purchase is
informative, even though purchases at the same inherited price pair
become uninformative after that update.

Let $(P_A(d),P_B(d))$ be the targets in Table~\ref{tab:targets}.
At belief $\mustar$, define the region and prescribed reset choices by
\begin{equation}
 \mathcal C_d=\{\mustar\}\times
 (\PP_d\cap[0.79,P_A(d)])\times(\PP_d\cap[0.99,P_B(d)]),
 \qquad p_i^{\mathrm{reset}}=P_i(d).
 \label{eq:region}
\end{equation}
No consumer compares anywhere in this region, and default purchases
have strictly positive surplus. Purchases leave the belief at $\mustar$;
the prescribed resets also keep the state in the region. If each firm
uses its target when it next receives an opportunity, the two installed
prices eventually become the targets.

\begin{table}[tb]
\centering\small\singlespacing
\caption{Equilibrium reset targets}\label{tab:targets}
\begin{tabular}{@{}lrrr@{}}
\toprule
Menu tick $d$ & $P_A(d)$ & $P_B(d)$ & $P_B(d)-P_A(d)$\\\midrule
$0.010$ & $0.910$ & $1.080$ & $0.170$\\
$0.005$ & $0.910$ & $1.085$ & $0.175$\\
$0.001$ & $0.914$ & $1.085$ & $0.171$\\
\bottomrule
\end{tabular}
\caption*{The central type law, hidden defaults, $\lambda=1$,
$\rho=0.25$, and $0<\alpha\le0.001$ apply to all three rows.
Targets are the highest menu prices below $R_A(\mustar)$ and
$R_B(\mustar)$ in the verified construction. The table does not assert
monotonicity of the gap as the menu is refined.}
\end{table}

\Needspace{12\baselineskip}
\begin{proposition}\label{prop:equilibrium}
Maintain the central primitives and hidden defaults. For each
$d\in\{0.01,0.005,0.001\}$ and every $0<\alpha\le0.001$, there is
a stationary Markov equilibrium on the state space reachable from
$(\mu_0,p_A,p_B)=(1/2,0.79,0.99)$ such that firms use the targets in
Table~\ref{tab:targets} throughout $\mathcal C_d$.

The event that the first arrival buys $B$ before either firm resets
has probability greater than $0.37164213$ and enters $\mathcal C_d$.
Conditional on that event, all subsequent purchases are uninformative,
and the installed prices eventually equal $(P_A(d),P_B(d))$ almost surely.
\end{proposition}

The probability in the proposition follows from competing independent
clocks. The unconditional probability of a $B$ purchase at the initial
state is $D_B=1/2-H/4=0.372385415\ldots$. The first event is an
arrival with probability $1/(1+2\alpha)$, giving an entry probability
$D_B/(1+2\alpha)$. Its smallest value in the stated range exceeds the
reported bound. Other histories may also enter a silent region, but are
not needed for this claim. In particular, the bound is not the full
probability of eventual informational silence.

\subsection{Why firms do not deviate}

Inside $\mathcal C_d$, each firm serves its default half of consumers.
Raising its price within the region increases revenue without changing
learning. This gives a simple reason for the candidate target: it is
the largest menu price below its default consumers' reservation value.
At $\mustar$ these values are approximately $0.914326$ for $A$ and
$1.085674$ for $B$. The target prices lie strictly below them, while
the next higher menu prices lie strictly above them.

This observation is not sufficient for equilibrium. A firm might cut
far enough to activate comparison, or raise its price and try to sell
to informed buyers rather than defaults. The verification therefore
divides all alternatives into three classes. A price within the silent
rectangle but below the target earns less until the firm's next own
reset, while leaving future beliefs and the rival's behavior unchanged.
A price above the target makes no sales before the next reset: at the
reservation boundary neither default group finds comparison worthwhile,
and increasing the deviating price further cannot induce it. The
remaining alternatives are lower prices outside the rectangle. Their
value must be bounded along the learning paths that they generate.

Voluntary participation provides a useful ceiling. Under any admissible
pricing and continuation behavior, a firm's normalized value cannot
exceed $R_i(\mu)$. A consumer purchases only when the price does not
exceed her conditional expected value, net of a nonnegative switching
cost. Taking expectations and using the posterior martingale property
gives the ceiling for the entire continuation. This argument does not
assume that the firm actually extracts that value.

To bound a lower-price deviation, keep the resulting pair fixed until
the first reset, whose rate is $h=2\alpha$. Let $F_{i,s}$ be its expected revenue per arrival averaged over time
with exponential discount rate $s=\rho+h$, normalized so that a
constant revenue stream $u$ gives $F_{i,s}=u$. Whatever happens after the reset, the deviation's normalized
value is at most
\begin{equation}
 \overline V_i=\frac{\rho F_{i,\rho+2\alpha}
                      +2\alpha R_i(\mustar)}{\rho+2\alpha}.
 \label{eq:deviation}
\end{equation}
The purchase tree is used to bound $F_{i,s}$, retaining valid upper
bounds on all unresolved continuations. Rigorous numerical bounds cover every relevant lower price for every
rival price in the rectangle, and each bound is strictly below $P_i(d)/2$.
Appendix~\ref{app:equilibrium} gives the argument and the largest bounds;
Appendix~\ref{app:computational} documents the implementation and reproduction. Thus the verification includes
the information generated by deviations, not just their immediate
effect on market shares.

The same deviation provides a direct comparison with the
independent-quality economy. At the cent targets $(0.91,1.08)$, a cut
by $B$ to $0.85$ then earns $a_0$ at every arrival until the next reset.
Counting only that revenue gives value
$\rho a_0/(\rho+2\alpha)>0.55726>0.54$ throughout the stated reset-rate
range. All later profits are nonnegative. Hence the cent target pair
cannot remain absorbing in this control, whatever continuation play is
selected. With persistent quality, the verified bounds reject the same
cut. Appendix~\ref{app:equilibrium} gives the comparison, which concerns
a pricing incentive at a matched state rather than the control economy's
entire equilibrium.

The final step is to supply equilibrium outside $\mathcal C_d$.
Temporarily restrict choices in the region to the prescribed targets.
The finite price and observation sets generate a countable state space
from the initial state. Discounted finite-state approximations have
stationary equilibria, and a diagonal limit supplies one for the
restricted game. Restoring all actions in $\mathcal C_d$ does not
create a profitable deviation, because the preceding bounds allow any
admissible continuation after an exit. The details are in
Appendix~\ref{app:equilibrium}. This establishes existence on the
reachable component without requiring a computed global reset policy.

\subsection{The extent of the equilibrium claim}

The construction selects an equilibrium with a positive-probability route
into silence. Conditional on entry, the belief stops changing immediately;
the target prices are installed only as each firm's next reset opportunity
arrives. Other equilibria and other initial conditions need not produce
this outcome.

The reset-rate restriction is substantial. At $\alpha=0.001$, a firm's
mean wait for its next reset is 1,000 arrival-time units, equivalent to
250 discount horizons at $\rho=0.25$. These values make the construction
an existence result, not a quantitative account of observed adjustment
frequencies. The shorter price-spell comparison in
Section~\ref{sec:incentive} establishes that the incentive can arise
much sooner, but does not overcome the continuation-value problem in
the full game. Equilibrium at materially faster reset rates would need
a sharper analysis of that continuation.

The two finer menus address a different restriction. On adding new
prices it is necessary to adjust the targets: the cent-menu pair cannot
be defended against every newly feasible increase. With the targets
in Table~\ref{tab:targets}, however, the verification succeeds on both
finer menus. This establishes the construction on the three specified menus.
Continuous prices would require a separate argument at inspection and
participation boundaries, as well as an equilibrium-existence argument.

There are two further comparisons worth separating from the main
construction. First, replacing independent reset clocks by a common
clock at the same individual frequency also supports the cent-menu
construction. Asynchronous adjustment therefore affects the transition
of installed prices, but is not necessary for the silent unequal-price
outcome. Second, eliminating the private signal while retaining costly
rival access supports the cent-menu targets in a fixed-belief economy.
Appendix~\ref{app:comparisons} gives both arguments. Price dispersion
alone cannot identify the role of learning; Proposition~\ref{prop:incentive}
provides the direct comparison of pricing incentives.

Finally, the cent-menu verification survives a small neighborhood of
the cost and value specification, and it has also been checked when
defaults are publicly observed. Those extensions, stated precisely in
Appendix~\ref{app:equilibrium}, establish local nondegeneracy rather
than an economically broad parameter range. More consequentially,
arbitrarily rare cheap comparison can destroy exact silence. We turn
to that possibility next.

\section{Rare inspection and the duration of inertia}\label{sec:duration}

\subsection{A population with rare cheap comparison}

The equilibrium construction uses a comparison-cost floor high enough
to make every consumer refrain from comparing in the relevant region.
In applications, some consumers may be able to compare more cheaply.
Exact silence is then a demanding outcome. It is useful to ask a
different question: when most purchases are uninformative, how long
does it take for the information generated by the remaining consumers
to move public beliefs appreciably?

For this section hold prices fixed at $(0.91,1.08)$, start at
$\mustar$, and let $I=[0.411,0.419]$. This is a new experiment at
specified prices, not an assertion that the equilibrium targets remain
optimal after the type distribution changes. Within each default
group, consider the consumers with $K<0.1$, whose share is
$H=1-e^{-5/7}$. Select a fraction $\varepsilon/H$ of them independently
of quality and assign them comparison cost $C=0.01$. Everyone else
has $C=0.3$. Thus cheap types form a share $\varepsilon$ of each
default group and of the entire population, where $0<\varepsilon\le H$.

The construction preserves the exponential marginal distribution of
switching costs. It deliberately correlates cheap comparison with low
switching costs, ensuring that a cheap consumer can act upon her signal.
The selection can be implemented by an independent private random
draw conditional on $K<0.1$. Types remain independent across arrivals
and of quality. The perturbation varies the mass of cheap types while keeping their
cost and the marginal distribution of $K$ fixed.

Throughout $I$, ordinary consumers continue to buy their defaults
without comparing. Every cheap consumer compares and buys the product
favored by her signal, whatever her default. Both comparison and the
favored purchase are strictly worthwhile on the entire interval.
These incentives are verified in Appendix~\ref{app:duration}.
Consequently, every arrival produces a purchase before the posterior
leaves $I$, and the observation law has a particularly simple form.

We compare two public records. With hidden defaults, an observer sees
only the product purchased. With observed defaults, she also sees the
buyer's initial default. Prices and consumer choice rules at any common belief in $I$ are the
same in both experiments. In particular, observed defaults do not allow
firms to personalize prices before the purchase, and the observer
does not directly see whether the buyer compared.

\subsection{Exit times and observability}

Let $\mu_t^H$ and $\mu_t^O$ be the public posterior under the two
observation regimes. Define the first exit times by
\begin{equation}
 \tau_r=\inf\{t\ge0:\mu_t^r\notin I\},\qquad r\in\{H,O\}.
 \label{eq:exit}
\end{equation}
The interval, prices, prior, and signal accuracy are held fixed as
$\varepsilon$ varies. Exit is an information event: it records a
departure from a specified neighborhood of the silent belief. It
does not record the date at which a firm decides to change its price.

\begin{samepage}
\Needspace{10\baselineskip}
\begin{proposition}\label{prop:duration}
For the type family and fixed prices described above, both observation
regimes are informative for every $\varepsilon>0$ before exit from
$I$. Starting at $\mustar$, expected exit times satisfy
\begin{equation}
 \E\tau_H=\Theta\!\left(\frac{1}{\lambda\varepsilon^2}\right),
 \qquad
 \E\tau_O=\Theta\!\left(\frac{1}{\lambda\varepsilon}\right)
 \quad\text{as }\varepsilon\downarrow0.
 \label{eq:orders}
\end{equation}
The constants in these bounds depend on the maintained primitives and
the interval, but not on $\varepsilon$. Expectations use the initial
prior $\mustar$.
\end{proposition}
\end{samepage}

The difference comes from how an informed decision can be recognized.
With hidden defaults, a purchase from $A$ may reflect either an
ordinary consumer's default or a cheap consumer's favorable signal.
The first group is large, while the second is rare. Let $a=2q-1=1/2$.
The probability of an $A$ purchase is $(1+a\varepsilon)/2$ in state
$A$ and $(1-a\varepsilon)/2$ in state $B$. Hence the likelihood
ratio of an individual purchase is close to one when $\varepsilon$
is small. Most observed purchases change beliefs only slightly, including
purchases made by consumers who did not themselves compare.

When defaults are observed, a switch can be separated from a purchase
at the default. Only cheap types switch in this experiment. Observing
an $A$-default consumer buy $B$ therefore reveals a negative signal,
and observing the reverse switch reveals a positive signal. Either
switch moves the posterior outside $I$. Switches occur with probability
$\varepsilon/2$ per arrival, independently of the quality state.
The wait for such an observation is therefore of order
$1/(\lambda\varepsilon)$.

The information-rate comparison is shown in
Figure~\ref{fig:information}. The vertical axis measures the
Kullback--Leibler divergence between the two quality-state distributions
of one observation, in nats per arrival. With hidden defaults this
divergence is of order $\varepsilon^2$; with observed defaults it is
of order $\varepsilon$. For example, at $\varepsilon=0.01$, the
divergences are approximately $0.0000500004$ and $0.00275909$.
Their ratio is about $55.18$. This is a ratio of information rates,
not a ratio of expected exit times.

\begin{figure}[tb]
\centering\singlespacing
\begin{tikzpicture}
\begin{loglogaxis}[width=.91\textwidth,height=7.1cm,
 xmin=.0001,xmax=.05,ymin=.000000003,ymax=.02,
 xlabel={Cheap-inspector share $\varepsilon$},
 ylabel={Information per arrival (nats)},
 tick label style={font=\small},label style={font=\small},
 grid=both,grid style={gray!15},minor grid style={gray!8},
 legend style={font=\small,at={(.03,.97)},anchor=north west,draw=none,fill=white}]
\addplot[firmblue,thick] coordinates {\HiddenInformationCoordinates};
\addlegendentry{Hidden defaults: order $\varepsilon^2$}
\addplot[learnred,thick,dashed] coordinates {\ObservedInformationCoordinates};
\addlegendentry{Observed defaults: order $\varepsilon$}
\end{loglogaxis}
\end{tikzpicture}
\caption{Rare inspection and the information in purchases. The curves use the
exact observation laws in Appendix~\ref{app:duration}, at $q=0.75$.
They apply while the posterior is in $I$ and the indicated consumer
choices hold. Prices are fixed at $(0.91,1.08)$. These are information
rates, not simulated durations of price or belief spells.}
\label{fig:information}
\end{figure}

Information rates provide intuition, but are not by themselves a proof
of the exit-time orders. For hidden defaults, the one-arrival posterior
variance is bounded by a constant times $\varepsilon^2$. A stopped
martingale argument then gives a lower bound on expected exit time.
For the upper bound, log odds follow a random walk with step size of
order $\varepsilon$ and state-conditional drift of order
$\varepsilon^2$. The bounded interval and controlled overshoot give
the matching upper order. With observed defaults, waiting for the
first switch supplies the upper bound, while a one-arrival variance
bound proportional to $\varepsilon$ supplies the lower bound.
Appendix~\ref{app:duration} gives all four inequalities.

\subsection{Long information spells without exact silence}

The duration result gives a useful sense in which the silent benchmark
can approximate an informative market. For the hidden-default
experiment at $\varepsilon=0.001$ and $\lambda=1$, the probability
that the posterior leaves $I$ within four time units is at most
$0.022596$. Four time units equal one discount horizon at the central
discount rate. Although every arrival is informative, the public belief
is therefore very likely to remain in this interval over that horizon.
The number is an upper bound for a specified experiment, not an estimate
of an observed market's adjustment frequency.

Revealing defaults does not matter for exact silence in the same way
that it matters for nearby informative states. Under the maintained
independence assumptions, the two default groups' responses to the
binary signal cannot offset each other: a signal favoring $A$ can
only raise its purchase probability within either group. If the
aggregate purchase experiment is exactly silent, revealing the default
does not restore information. Appendix~\ref{app:consumer} proves
this statement, allowing the outside option. Once cheap consumers
respond to signals, however, defaults help identify which purchases
contain information. Thus equality at the silent benchmark is
consistent with very different learning times close to it.

The comparison concerns the limit as the cheap share vanishes. It does
not order expected exit times at every positive share, or cover every
way of assigning cheap costs. In particular, cheap comparison need not
make purchases responsive if the selected consumers face prohibitive
switching costs. The stated family links low comparison costs to buyers
who can act on their signals.

The result establishes exit from $I$, not eventual revelation of quality.
Here the cheap cost remains $0.01$ as its population share falls. A pricing
equilibrium under this perturbation would also have to allow firms to
react before that exit. The observation laws and duration bounds provide
the information component of such an extension.

\section{Conclusion}\label{sec:conclusion}

When comparison is costly, a price cut changes more than current
market shares. It can induce consumers to obtain information and make
their purchases informative to later buyers. If the lower price remains
installed, the resulting change in future demand can overturn the
current sales gain. The three-arrival comparison identifies this
incentive directly: the specified cut loses with persistent quality
but continues to pay when the current choice problem is repeated with
independent quality draws.

The equilibrium construction establishes that silent unequal prices
can be reached after an informative purchase and supported against
all available deviations on the specified finite menus. Its slow-reset
restriction remains substantial. The role of the construction is to
make the phenomenon nonvacuous and the pricing choices explicit, rather
than to give a general explanation of price dispersion. Costly access
can generate a price gap without learning, and asynchronous adjustment
is not necessary for the constructed outcome.

Rare inspection replaces exact silence with a question about duration.
In the fixed-price experiment, observing defaults changes the order
of the time needed to leave a quiet belief interval because it makes
rare switches identifiable. Two extensions would further connect this
information result to pricing: establishing equilibrium with continuous
prices, and solving firms' responses to rare comparison at less
restrictive reset rates. Both require additional equilibrium analysis.
The present results isolate the incentive and the observation mechanism
which that analysis would need to preserve.
\label{maintextend}

\clearpage
\appendix
\setstretch{1.15}
\numberwithin{equation}{section}
\numberwithin{table}{section}

\section{Consumer choice and the information in purchases}\label{app:consumer}

\subsection{A kernel with an outside option}

The signal probabilities and posteriors at public belief $\mu$ are
\[
 \pi_+=\tfrac14+\tfrac12\mu,\quad \pi_-=1-\pi_+,
 \qquad \mu_+=\frac{3\mu}{1+2\mu},\quad
 \mu_-=\frac{\mu}{3-2\mu}.
\]
The formulas at degenerate priors are interpreted by continuity. For a
consumer with default $j$ and rival $k$, let
\[
 \begin{aligned}
 u_{js}&=R_j(\mu_s)+Z-p_j,&
 b_j&=\max\{R_j(\mu)+Z-p_j,0\},\\
 L_{js}&=\max\{u_{js},0\},&
 h_{js}&=R_k(\mu_s)+Z-p_k-L_{js}.
 \end{aligned}
\]
Here $Z=0$ in the central economy. The small common value shift $Z$ is
used only in the local extension below. After paying $C$, expected gross
surplus is $\sum_s\pi_s[L_{js}+(h_{js}-K)_+]$. Thus comparison occurs
precisely when
\begin{equation}
 \sum_s\pi_s(h_{js}-K)_+>t_j,
 \qquad t_j=C+b_j-\sum_s\pi_s L_{js}.
 \label{eq:cutoffcondition}
\end{equation}
This formulation retains participation both before and after the signal.

For the central comparison cost and its stated small perturbations,
$t_j$ is strictly positive. To see why, write $\Delta=v_H-v_L$ and
consider a consumer who would exit after one of the signals. The
largest expected surplus available after only the other signal is
bounded by
\[
 \chi(q)=\Delta\frac{\sqrt q-\sqrt{1-q}}
                         {\sqrt q+\sqrt{1-q}}.
\]
Indeed, if every purchase utility is nonpositive after signal $s$,
each utility after the other signal $s'$ is at most
$\Delta|\mu_+-\mu_-|$: the absolute slope of either product's value
in the posterior is $\Delta$. Expected gross surplus is therefore at
most $\pi_{s'}\Delta|\mu_+-\mu_-|$. Maximizing this expression over
$\mu$ and $s'$ gives $\chi(q)$. For example,
$\pi_+(\mu_+-\mu_-)=
(2q-1)\mu(1-\mu)/(q-(2q-1)\mu)$, whose maximum is the displayed
bound; the other case follows by symmetry.
At $q=3/4$ and $\Delta=1$, the bound is $2-\sqrt3<0.268$.
The value of information about the default purchase alone is no larger,
so $t_j\ge C-\chi(q)>0$ when $C\ge0.295$.
The same bound shows that a consumer who compares cannot optimally exit
after either signal: if all purchase utilities were nonpositive after
one signal, expected surplus after the other could not cover $C$.
This argument is independent of posted prices and of a common value shift.

Since the left side of \eqref{eq:cutoffcondition} is the maximum of
three decreasing affine expressions and zero, the comparison cutoff is
\begin{equation}
 k_j=\max\left\{h_{j+}-\frac{t_j}{\pi_+},\quad
 h_{j-}-\frac{t_j}{\pi_-},\quad
 \pi_+h_{j+}+\pi_-h_{j-}-t_j\right\}.
 \label{eq:kernelcutoff}
\end{equation}
A consumer compares if $K<k_j$ and switches after signal $s$ if also
$K<h_{js}$. Let $H(x)=1-\exp(-[x]_+/0.14)$,
$H_j=H(k_j)$, $S_{js}=H(\min\{k_j,h_{js}\})$, and
$I_j=\1\{R_j(\mu)-p_j\ge0\}$. For the central law, conditional on
signal $s$ and default $j$, the probabilities of buying the default,
buying the rival, and exiting are, respectively,
\begin{equation}
 I_j(1-H_j)+H_j-S_{js},\qquad S_{js},\qquad
 (1-I_j)(1-H_j).
 \label{eq:choices}
\end{equation}
These probabilities include both consumers who compare and those who do
not. The fictitious conditioning on a signal for a non-comparer merely
uses its independent distribution; it does not assume she observes it.

Average \eqref{eq:choices} over the two equally likely defaults to obtain
$g_s(y)$, for $y\in\{A,B,0\}$. The state-conditional observation kernel is
\begin{equation}
 \ell_A(y)=qg_+(y)+(1-q)g_-(y),\qquad
 \ell_B(y)=(1-q)g_+(y)+qg_-(y).
 \label{eq:likelihoodkernel}
\end{equation}
Together with Bayes' rule in the main text, this gives demand and the
next public belief at every state. In particular, non-purchase has the
same likelihood in the two states: all comparers buy, while the decision
not to compare is made before the signal. A no-sale arrival therefore
leaves the posterior unchanged. When defaults are observed, retain the
two default-specific kernels instead of averaging them.

For correlated $(C,Z,K)$, integrate the underlying choice indicators
over the joint law. One cannot insert a conditional exponential
switching-cost distribution unless that conditional distribution is an
assumption. The uniform calculations below use marginal bounds instead.
For example, if all relevant type-specific cutoffs lie between $a$ and
$b$, the comparison probability lies between $\Pp(K<a)$ and
$\Pp(K\le b)$. This remains valid with arbitrary correlation.
For costs as low as $0.01$, the preceding no-exit argument is not used;
Appendix~\ref{app:duration} verifies the choices directly.

\subsection{Silence in the covered subproblem}

The following support characterization is useful for understanding the
role of cost floors. It applies to states at which the outside option
never binds, including after comparison; it is not a global demand
formula for the participation model. For this subsection, allow $q\in(1/2,1)$, $\Delta>0$,
$\mu\in(0,1)$, default share $\phi\in(0,1)$ for $A$, and joint
laws of nonnegative $(C,K)$ within defaults. Replace the numerical
signal formulas above by
$\pi_+=(1-q)+(2q-1)\mu$, $\pi_-=1-\pi_+$,
$\mu_+=q\mu/\pi_+$, and $\mu_-=(1-q)\mu/\pi_-$.
Put
\[
 g=p_A-p_B,\qquad v_s=\Delta(2\mu_s-1),\qquad w=v_+-v_-.
\]
A default-$A$ consumer changes her purchase with the signal exactly on
\[
 \mathcal R_A=\{g-v_+\le K<g-v_-,\quad
                    C<\pi_-(g-v_--K)\};
\]
a default-$B$ consumer does so on
\[
 \mathcal R_B=\{v_--g\le K<v_+-g,\quad
                    C<\pi_+(v_+-g-K)\}.
\]
Let $r_j$ be the conditional probability of $\mathcal R_j$ in default
group $j$. Directly separating always-default, always-switch, and
signal-responsive choices gives
\begin{equation}
 \ell_A(A)-\ell_B(A)=(2q-1)\{\phi r_A+(1-\phi)r_B\}.
 \label{eq:responsive}
\end{equation}
All terms are nonnegative. Silence is therefore equivalent to
$r_A=r_B=0$, rather than cancellation of informative responses.

At $\mu=1/2$ and equal prices, write $\gamma=(2q-1)\Delta$.
The condition becomes
\[
 \Pp(K+2C<\gamma\mid J=j)=0,\qquad j=A,B.
\]
It is a joint-support condition. If $C$ and $K$ are independent within
each group and have lower support endpoints $\underline c$ and $\kappa$,
with positive mass arbitrarily near those endpoints, it reduces to
$2\underline c+\kappa\ge\gamma$. Marginal endpoints alone are
insufficient under correlation.
For common $C=c$ and switching costs with continuous positive density
on $(\kappa,\infty)$, define
\[
 \Psi_A=\pi_-\min\{w,[g-v_--\kappa]_+\},\qquad
 \Psi_B=\pi_+\min\{w,[v_+-g-\kappa]_+\}.
\]
Silence is equivalent to $\Psi_A\le c$ and $\Psi_B\le c$.
These inequalities describe a possibly empty interval of silent price
gaps: its lower endpoint is $v_+-\kappa-c/\pi_+$ if
$c<\pi_+w$, and $-\infty$ otherwise; its upper endpoint is
$v_-+\kappa+c/\pi_-$ if $c<\pi_-w$, and $+\infty$ otherwise.
This geometry explains why both prices and the support of comparison
costs matter, subject to the maintained coverage restriction.

\subsection{Revealing the default}

The conclusion about exact silence also holds with an outside option.
For every type, raising the signal from $-$ to $+$ weakly raises the
chance of an $A$ purchase and weakly lowers that of a $B$ purchase.
Comparison is chosen before the signal. Since types and defaults are
independent of quality, the state-likelihood differences of an $A$
purchase are nonnegative within each default group; those of a $B$
purchase are nonpositive. Equality in the aggregate therefore forces
equality in each group for both purchases, and hence for non-purchase.
The converse follows by aggregation. Thus hidden-default silence and
observed-default silence coincide under these independence assumptions.
Away from silence, the law of total variance implies that observing
$(J,Y)$ weakly increases the variance of the next posterior relative to
observing $Y$. This information comparison does not order equilibrium
profits or preferred prices.

\section{The price-cut experiment}\label{app:incentive}

\subsection{The complete three-arrival tree}

Apply \eqref{eq:choices}--\eqref{eq:likelihoodkernel} at
$(\mustar,0.91,0.85)$. Table~\ref{tab:tree} gives every history needed
for the first three expected revenues. A row's mass is its unconditional
probability, and its revenue is the expected revenue at the next arrival.
The empty history is denoted by $\varnothing$; $0$ denotes no purchase.
All displayed values are rounded. The reproduction files retain outward
bounds for these quantities; Appendix~\ref{app:computational} describes their
precision and verification.

\begin{table}[htbp]
\centering\small
\caption{Purchase histories following the cut}\label{tab:tree}
\begin{tabular}{@{}lrrr@{}}
\toprule
History & Mass & Posterior $\Pp(\theta=A)$ & Next revenue of $B$\\\midrule
$\varnothing$ & $1.000000000$ & $0.414326273$ & $0.561718089$\\
$A$ & $0.339155189$ & $0.520326666$ & $0.425000000$\\
$B$ & $0.660844811$ & $0.359925327$ & $0.571659061$\\\midrule
$AA$ & $0.169577595$ & $0.520326666$ & $0.425000000$\\
$AB$ & $0.169577595$ & $0.520326666$ & $0.425000000$\\
$BA$ & $0.086003653$ & $0.627831247$ & $0.322606949$\\
$BB$ & $0.444444616$ & $0.308083349$ & $0.576125038$\\
$B0$ & $0.130396541$ & $0.359925327$ & $0.571659061$\\
\bottomrule
\end{tabular}
\end{table}

Multiplying each next-arrival revenue by its history mass and summing
at each depth gives the three $a_n$ in Proposition~\ref{prop:incentive}.
In particular, the $B0$ row contributes to $a_2$ even though it leaves
beliefs unchanged. With $g_n=a_n-0.54$, the unique zero of the
three-arrival gain is
\[
 \delta^*=\frac{-g_1-\sqrt{g_1^2-4g_2g_0}}{2g_2}
          =0.557128884106174\ldots.
\]
The signs $g_0>0$, $g_1,g_2<0$, and
$g_0+g_1=0.003636968435884\ldots>0$ establish both horizon claims.
In the independent-quality control each new consumer begins at
$\mustar$, so each arrival has revenue $a_0$ after the cut.
For a common affine terminal payoff $W(\mu)=a+b\mu$, the expected
terminal value is $W(\mustar)$ under either choice by the posterior
martingale property. Its discount factor is the same as well, since
arrival times are independent of quality and choices. Such a terminal
payoff therefore cancels. No corresponding conclusion follows for an
arbitrary nonlinear $W$.

The adverse expected continuation is not a global revenue
supermartingale property. For example, after history $BB$ the conditional
expected next-arrival revenue rises by more than $0.00605$. The proof
uses the explicitly averaged tree, not a sign asserted at every node.

\subsection{Uniform cost checks}

The homogeneous-cost calculation partitions $[0.295,0.305]$ into 100
closed intervals of width $0.0001$. For each whole interval, outward
arithmetic applied to the consumer kernel establishes silence at the
stay prices, a strictly positive initial gain from the cut, and a
strictly negative three-arrival gain for $\delta\in[0.8,1]$.
Every cell has $g_1<0$ and $g_2<0$, and the gain at $\delta=0.8$ is
negative. The quadratic gain therefore decreases on $[0.8,1]$.
The calculation encloses each whole cost interval, rather than evaluating
100 individual cost values. The initial belief remains
$\mustar$ throughout this exercise.

For arbitrary heterogeneity supported on $C\in[0.299,0.301]$, the
same switching-cost marginal, and arbitrary correlation between the two
costs, use upper and lower cutoff envelopes as in
Appendix~\ref{app:consumer}. The interval calculation gives initial cut revenue
above $0.56047319$ and three-arrival weighted average revenue below
$0.53740574$ at $\delta=0.8$, while staying gives $0.54$.
The relevant posterior intervals are propagated through every branch.
These bounds hold for all joint laws on the stated support, but do not
extend to arbitrary heterogeneity on the wider homogeneous-cost interval.
Neither exercise asserts an equilibrium or reconstructs the entry belief
for every altered population.

\subsection{An independently ending price spell}

Let a spell end at an independent exponential time of rate $h$, put
$s=\rho+h$, and keep the cut pair installed until that time. Define
\[
 F_s=s\int_0^\infty e^{-st}\E[u_B(\mu_t)]\,dt,
\]
where $u_B(\mu)$ is expected revenue per arrival at the fixed pair.
The expected normalized revenue before the spell ends is
$(\rho/s)F_s$. Staying instead gives $(\rho/s)b$.
For $\rho=0.25$ and $\lambda=1$, the saved bounds are
\[
 \begin{array}{c|c|c}
 h & s & F_s\\\hline
 0.75 &1&[0.5364776000949237,\ 0.5364776015572333]\\
 1.75 &2&[0.5460561979677883,\ 0.5460561980100186].
 \end{array}
\]
Thus the cut loses in the first spell comparison and pays in the second.
The discounted-tree recursion and valid unresolved-tail bounds are the
same as those described in Appendix~\ref{app:equilibrium}.
If the end of the spell delivers a common normalized continuation
$W(\mu)$, the total gain is
\[
 \frac{\rho}{s}(F_s-b)+h\int_0^\infty e^{-st}
             \{\E[W(\mu_t)]-W(\mustar)\}\,dt.
\]
The last term vanishes for affine $W$ and is nonnegative for convex $W$.
These observations explain both the within-spell comparison and why it
cannot establish equilibrium for arbitrary post-reset continuation.

\section{Verification of reset incentives and equilibrium completion}\label{app:equilibrium}

\subsection{Silence, participation, and the entry event}

Inside the rectangle in \eqref{eq:region}, default surplus is strictly
positive. Its smallest values, attained at the target prices, are
\[
 \begin{array}{c|cc}
 d & R_A(\mustar)-P_A & R_B(\mustar)-P_B\\\hline
 0.010 &0.0043262732\ldots&0.0056737267\ldots\\
 0.005 &0.0043262732\ldots&0.0006737267\ldots\\
 0.001 &0.0003262732\ldots&0.0006737267\ldots.
 \end{array}
\]
As long as this surplus is nonnegative, comparison becomes weakly more
attractive when the default price rises or the rival price falls.
It is therefore sufficient to check each default group's highest own
price and lowest rival price. Formula~\eqref{eq:kernelcutoff} gives
negative switching-cost cutoffs at these corners. Upper bounds for
$(k_A,k_B)$ are $(-0.01428278,-0.00677311)$ on the cent menu,
$(-0.01428278,-0.00177311)$ on the half-cent menu, and
$(-0.01028278,-0.00177311)$ on the tenth-cent menu.
Since $K\ge0$, nobody compares anywhere in the rectangle.

At $(1/2,0.79,0.99)$, direct choice gives no comparison for default $A$
and comparison for default $B$ exactly when $K<0.1$. Such comparers
follow their signals and always buy. The likelihoods and Bayes update
in \eqref{eq:entry} follow. The initial purchase probability is
$D_B=1/2-H/4$, so the probability of this purchase before either reset
is $D_B/(1+2\alpha)>0.37164213$ for $\alpha\le0.001$.
After entry, silence and prescribed target resets preserve the rectangle.
Each positive-rate own reset clock rings in finite time almost surely,
which proves the conditional convergence of installed prices.

\subsection{An upper bound valid after any deviation}

A purchaser's willingness to pay does not exceed her conditional
expected product value, since comparison and switching costs are
nonnegative. Conditional on her full decision information, multiply
this inequality by the indicator of purchase and then take expectations.
Product values are nonnegative, so expected revenue from firm $i$ at
one arrival is at most $R_i(\mu)$. The same inequality holds under
arbitrary feasible future price choices. Applying it at each future
arrival and using the posterior martingale property yields
\begin{equation}
 V_i(\mu,p_A,p_B)\le R_i(\mu).
 \label{eq:valueceiling}
\end{equation}
The bounded values also justify optional sampling and the infinite
horizon limit by dominated convergence. In the extension with
$|Z|\le\eta$, replace the right side by $R_i(\mu)+\eta$.

For a fixed pair, let $w_y(\mu)$ be the unconditional observation
probability, $T_y(\mu)$ the next belief, and $u_i=p_iw_i$.
With $\beta=\lambda/(\lambda+s)$, first-arrival decomposition gives
\begin{equation}
 F_{i,s}(\mu)=(1-\beta)u_i(\mu)
                   +\beta\sum_y w_y(\mu)F_{i,s}(T_y(\mu)).
 \label{eq:fixedrecursion}
\end{equation}
This normalization follows from
$F_{i,s}=s\int e^{-st}\E[u_i(\mu_t)]dt$; the flow $u_i$ persists
until the next arrival. Because $T_0(\mu)=\mu$, one may collect the
no-sale term and divide by $1-\beta w_0(\mu)$. Iterating this recursion
with lower and upper bounds on unresolved nodes bounds the entire fixed
price spell, including arbitrarily many no-sale arrivals.

At every unresolved node, zero is a lower bound and
$\min\{p_i,R_i(\mu)\}$ is an upper bound. The finer-menu calculation
also uses a concave majorant $M_i(\mu)$ of instantaneous revenue.
For every belief bin, interval arithmetic bounds revenue over the whole
bin. Give each endpoint a height at least as large as the upper bounds
on its adjacent bins, and construct the upper concave hull of these
points using exact rational operations. This piecewise affine concave
function dominates revenue throughout every bin. The posterior
martingale property and Jensen's inequality imply
$F_{i,s}(\mu)\le M_i(\mu)$. Thus the smaller of this majorant, price,
and expected value is a valid upper bound at a pruned node. No branch
is assigned zero upper continuation merely because it is small or deep.

The calculation also retains the constraint that branch probabilities
sum to one. If branch $y$ has probability in $[\underline w_y,\overline
w_y]$ and payoff upper bound $v_y$, then, for any constant $c$,
\[
 \sum_yw_yv_y\le c+\sum_y\bigl\{(v_y-c)_+\overline w_y
                          +\min(v_y-c,0)\underline w_y\bigr\}.
\]
Taking the smallest such bound over the branch payoff bounds tightens
interval averaging without assuming independence of probabilities and
continuations. For a no-sale self-loop, start with a valid upper value
and repeatedly apply this monotone upper-bound recursion; each iterate
remains valid. The option in the code that imposes default participation
is used only at roots where both default surpluses are known to be
nonnegative, never at descendant beliefs. On the verified lower-price
blocks and rival rows this follows directly from the price rectangle.

Let $T$ be the first reset after a deviation, with rate $h=2\alpha$.
Its normalized expected continuation is bounded using
\eqref{eq:valueceiling}. Independence of $T$ and the posterior
martingale give
\[
 V_i^{\mathrm{dev}}\le
 B_i(h):=\frac{\rho F_{i,\rho+h}(\mustar)+hR_i(\mustar)}{\rho+h}.
\]
It suffices to check the largest claimed reset rate. Indeed,
\begin{equation}
 B_i(h)=R_i(\mustar)-\rho\int_0^\infty e^{-(\rho+h)t}
              \{R_i(\mustar)-\E[u_i(\mu_t)]\}\,dt
 \label{eq:hazardmonotonicity}
\end{equation}
is increasing in $h$, since the expression in braces is nonnegative.
This monotonicity applies to the conservative bound, without asserting
monotonicity of the equilibrium value itself.

\subsection{Reset incentives}

Consider a reset state in $\mathcal C_d$. Choosing a lower own price
$p$ still inside the rectangle leaves beliefs and the rival's prescribed
behavior unchanged. Until the next own reset revenue is $p/2$, and
after choosing the target at that reset it is $P_i/2$. The value is
\[
 \frac{\rho p+\alpha P_i}{2(\rho+\alpha)}<\frac{P_i}{2}
 \qquad(p<P_i).
\]
This comparison uses the own reset hazard $\alpha$, not the hazard of
either firm's reset.

For prices above the target, the next tick already exceeds the default
reservation value. At that reservation boundary both default groups have
negative comparison cutoffs, for every rival price in the rectangle.
For example, on the finer menus, upper bounds at the worst rival corners
for the two groups are $(-0.00995651,-0.22109938)$ when $A$ charges
$R_A(\mustar)$, and $(-0.00109938,-0.22995651)$ when $B$ charges
$R_B(\mustar)$. The cent-menu rival range is contained in these ranges.
Above the reservation value, the default's outside option is already
zero and raising this price can only reduce the attractiveness of
comparison for either group. The deviator makes no sales until the next
reset. Its value is at most $2\alpha R_i(\mustar)/(\rho+2\alpha)$,
which is strictly below $P_i/2$ at the stated rates.

It remains to consider all prices below the rectangle. For each rival
menu price inside it, interval arithmetic gives a strict upper bound
$B_i(2\alpha)<P_i/2$. Table~\ref{tab:bounds} reports the largest upper
bounds, rounded upwards. The cent-menu calculation allows the larger
hidden-default rate $0.0025$, so it also covers the main result's rate
$0.001$. The finer-menu calculation is at $0.001$.

\begin{table}[htbp]
\centering\small
\caption{Upper bounds on the values of lower-price deviations}\label{tab:bounds}
\begin{tabular}{@{}lrrrr@{}}
\toprule
Menu & Upper bound, $A$ & Stay, $A$ & Upper bound, $B$ & Stay, $B$\\\midrule
Cent & $0.449300516$ & $0.455$ & $0.539587133$ & $0.540$\\
Half cent & $0.454764252$ & $0.455$ & $0.542396377$ & $0.5425$\\
Tenth cent & $0.456986832$ & $0.457$ & $0.542497607$ & $0.5425$\\
\bottomrule
\end{tabular}
\end{table}

Appendix~\ref{app:computational} documents the complete price coverage,
outward rounding, and checks on the saved bounds. The smallest reported
margin is about $2.39\times10^{-6}$. The equilibrium conclusion uses
these strict inequalities for every required menu choice.

\subsection{The independent-quality reset comparison}

Consider the cent targets and replace persistent quality by independent
quality draws at each arrival, with probability $\mustar$ of state $A$.
Keep costs, signals, prices, the cent menu, and reset rates unchanged.
This matches the consumer problem at the target state; it is not a
claim that this control reaches that state from prior $1/2$.
If the target pair remained absorbing, $B$ would receive value $0.54$.
After a cut to $0.85$, every arrival until the next reset instead gives
expected revenue $a_0$. Ignoring the nonnegative continuation yields
\[
 V_B^{\mathrm{cut}}\ge\frac{\rho a_0}{\rho+2\alpha}
                  >0.55726000>0.54,
 \qquad 0<\alpha\le0.001.
\]
The inequality already follows from the cent-menu calculation's
conservative lower bound $a_0>0.5617180887765$; the central-law value is
slightly higher. Consequently no equilibrium in this state-matched
control can keep the cent target pair absorbing. Under persistent
quality, the all-deviation verification rejects the same cut. This is a
comparison of reset incentives, not a ranking of the two economies'
equilibrium prices or welfare.

\subsection{Completing strategies outside the rectangle}

Temporarily restrict each firm's reset choices inside $\mathcal C_d$ to
its target. Uniformize arrivals and resets at rate
$\Lambda=\lambda+2\alpha$. The resulting discrete-event stochastic
game has discount factor $\Lambda/(\Lambda+\rho)<1$, bounded rewards,
finite actions, and finite transition support at each state-action pair.
One can include the identity of the next event in the state; only the
firm receiving a reset moves at that event. From the initial state,
closure under all feasible actions and positive-probability observations
therefore produces a countable state space.

Take the states reachable in at most $N$ events, and terminate at a
zero-payoff absorbing state upon leaving this finite set. Each such
finite discounted stochastic game has a stationary mixed equilibrium.
For completeness, at a candidate stationary profile compute its unique
discounted value vector. At each state let the acting player's allowed
mixed action be supported on maximizers of current reward plus the
discounted candidate continuation. This defines a nonempty, convex-valued,
upper hemicontinuous correspondence on the finite product of mixed-action
simplexes. Continuity of discounted values and Kakutani's theorem give
a fixed point, whose one-step inequalities imply optimality against all
strategies. This is the finite-game step underlying the usual discounted
countable-state approach; see also \citet{Federgruen1978}.

Enumerate the countable state space. Compactness of each finite-action
simplex gives a diagonal subsequence of the finite-game policies that
converges at every state. For a fixed starting state and finite horizon,
only finitely many successors matter, and all are retained for large
$N$. Finite-horizon payoffs thus converge under these policies. Bounded
rewards and a common discount factor make the omitted discounted tail
uniformly negligible. Each one-step best-response inequality passes to
the limit. The limiting stationary profile is consequently an equilibrium
of the restricted game on the entire reachable component.

Restore the excluded actions inside $\mathcal C_d$. Lower choices in
the rectangle lose revenue until the next own reset. All other choices
are strictly worse even under the continuation ceiling that permits
arbitrary feasible behavior after the first reset. Hence every restored
one-step deviation is unprofitable. The discounted one-step deviation
principle gives optimality against arbitrary continuation deviations,
proving Proposition~\ref{prop:equilibrium}. This argument selects some
equilibrium outside the region; it neither computes that policy nor
establishes uniqueness.

\subsection{A local cent-menu extension}

A separate interval calculation permits an initial prior satisfying
$|\mu_0-1/2|\le10^{-12}$ and private types with
\[
 |C-0.3|\le10^{-12},\qquad |Z|\le10^{-12},\qquad
 K=\kappa+W,\qquad 0\le\kappa\le10^{-12},
\]
where $W$ has an exponential marginal of mean $0.14$ within each default
group. The variables $C,Z,W$ may be correlated; types remain independent
across buyers and of quality. The shift $\kappa$ is a common parameter.
For the cent menu and inherited prices $(0.79,0.99)$, the first specified
$B$ purchase has a posterior between $0.414326273244525$ and
$0.414326273273533$. It lies in the verified strip
$|\mu-\mustar|\le10^{-10}$.

The target policy is silent throughout that strip and the cent-price
rectangle. Cutoff envelopes and the continuation ceiling
$R_i(\mu)+10^{-12}$ yield strict deviation inequalities uniformly over
the stated joint laws. Hidden defaults permit $\alpha\le0.0025$.
Observed defaults permit $\alpha\le0.001$, with upper deviation values
below $0.453685130$ and $0.538236937$ for $A$ and $B$, respectively.
Revealing defaults leaves exact silence unchanged, but the informative
deviation trees are checked separately; an information ordering alone
would not establish those incentive bounds. Equilibrium completion uses
the same finite-action argument. These extremely small neighborhoods
show local nondegeneracy. They are not a broad robustness range and do
not transfer to the two finer menus without a new verification.

\section{Rare inspection and exit times}\label{app:duration}

\subsection{Choice rules on the belief interval}

Let $I=[\underline\mu,\overline\mu]=[0.411,0.419]$ and use the
population in Section~\ref{sec:duration}. Ordinary types buy their
defaults without comparing throughout $I$ at $(0.91,1.08)$.
For cheap types, direct comparison of all three post-signal options,
for both defaults and every $K\in[0,0.1]$, gives a purchase of $A$
after $+$ and $B$ after $-$. To verify these inequalities, use
$\mu_+=3\mu/(1+2\mu)$ and $\mu_-=\mu/(3-2\mu)$: the favored
product's surplus, even if switching costs $0.1$, exceeds $0.15$
after $+$ and $0.12$ after $-$, and its advantage over the other
product remains strictly positive throughout the interval.
The expected gain from comparing is minimized at $K=0.1$.
Substitution in the direct surplus comparison over the whole interval
gives net gains above $0.1707049$ for default $A$ and $0.1792049$
for default $B$, after the comparison cost $0.01$.
Thus every cheap type compares, follows her signal, and purchases.
This check uses actual choices, including participation, rather than
the high-cost no-exit argument of Appendix~\ref{app:consumer}.

\subsection{Observation laws and information rates}

Write $a=2q-1=1/2$. Before exit, hidden-default purchase probabilities are
\[
 \ell_A(A)=\tfrac12(1+a\varepsilon),\qquad
 \ell_B(A)=\tfrac12(1-a\varepsilon),
\]
with the complementary probabilities for $B$. For observed defaults,
the probabilities of $(A,A),(A,B),(B,A),(B,B)$ under state $A$ are
\[
 \tfrac12\bigl(1-\varepsilon(1-q),\ \varepsilon(1-q),\
                  \varepsilon q,\ 1-\varepsilon q\bigr).
\]
Under state $B$, replace $q$ by $1-q$. The two kernels differ for
all $\varepsilon>0$. Their state-$A$ Kullback--Leibler divergences
relative to state $B$ are
\begin{align}
 D_H&=a\varepsilon\log\frac{1+a\varepsilon}{1-a\varepsilon},
       \label{eq:klhidden}\\
 D_O&=\frac{a\varepsilon}{2}\left\{\log\frac q{1-q}
       +\log\frac{1-\varepsilon(1-q)}{1-\varepsilon q}\right\}.
       \label{eq:klobserved}
\end{align}
Expansion at zero gives $D_H=2a^2\varepsilon^2+O(\varepsilon^4)$ and
$D_O=\frac a2\log(q/(1-q))\varepsilon+O(\varepsilon^2)$.
Figure~\ref{fig:information} plots the exact expressions, not their
asymptotic approximations. At $\varepsilon=0.01$ the values are
$0.0000500004166729\ldots$ and $0.0027590935621770\ldots$.

\subsection{Hidden defaults: both exit-time bounds}

Put $d_I=\min\{\mustar-\underline\mu,\overline\mu-\mustar\}>0$.
Bayes' rule gives the one-arrival conditional variance of the posterior
increment:
\begin{equation}
 v_H(\mu)=\frac{4a^2\varepsilon^2\mu^2(1-\mu)^2}
                   {1-a^2\varepsilon^2(2\mu-1)^2}
 \le\frac{\varepsilon^2}{16(1-\varepsilon^2/4)}=:\bar v_H.
 \label{eq:hiddenvariance}
\end{equation}
The stopped posterior is a bounded martingale. Its expected squared
displacement at $T\wedge\tau_H$ is at most
$\lambda\bar v_H\E[T\wedge\tau_H]$. An exit entails displacement
of at least $d_I$, including an overshoot. Consequently,
\begin{equation}
 \Pp(\tau_H\le T)\le\frac{\lambda T\bar v_H}{d_I^2},
 \qquad \E\tau_H\ge\frac{d_I^2}{\lambda\bar v_H}.
 \label{eq:hiddenlower}
\end{equation}
For the expectation inequality, first stop at $T$, and then let $T$
increase. Almost-sure exit and integrability follow from the upper-bound
argument below. At $\varepsilon=0.001$, $\lambda=1$, and $T=4$, the
probability bound is less than $0.022595621$.

For the upper bound, let $L(x)=\log(x/(1-x))$ and
$\ell=\log((1+a\varepsilon)/(1-a\varepsilon))$. Until exit, log odds
move by $+\ell$ or $-\ell$ at each arrival. Under state $A$, the
probability of the positive step is $(1+a\varepsilon)/2$, so the mean
increment is $a\varepsilon\ell$. Under state $B$ its sign is reversed.
For each fixed $\varepsilon>0$, a sufficiently long consecutive run of
either purchase exits the bounded log-odds interval. Independent blocks
of that length give a geometric bound on the number of arrivals to exit,
so its expectation is finite. Wald's identity and an overshoot of at
most $\ell$ then give
\begin{align}
 \E[\tau_H\mid A]&\le
 \frac{L(\overline\mu)-L(\mustar)+\ell}{\lambda a\varepsilon\ell},
 \\
 \E[\tau_H\mid B]&\le
 \frac{L(\mustar)-L(\underline\mu)+\ell}{\lambda a\varepsilon\ell}.
 \label{eq:hiddenupper}
\end{align}
Since $\ell\sim2a\varepsilon$, these are of order
$1/(\lambda\varepsilon^2)$. Averaging over the initial prior and
combining with \eqref{eq:hiddenlower} proves the hidden-default claim.

\subsection{Observed defaults: both exit-time bounds}

A switch occurs with probability $\varepsilon/2$ per arrival in either
quality state. An $A$-default buyer switching to $B$ reveals a negative
signal; a $B$-default buyer switching to $A$ reveals a positive signal.
The resulting posteriors are $\mu_-$ and $\mu_+$, respectively, both
outside $I$ for every $\mu\in I$. Thus exit occurs no later than the
first switch, if it has not occurred already. This gives
\[
 \E\tau_O\le \frac{2}{\lambda\varepsilon}.
\]
To get the reverse order, consider additionally revealing whether the
consumer is cheap. This indicator is independent of quality and hence
uninformative on its own. Conditional on an ordinary type, the observed
default purchase contains no information. Conditional on a cheap type,
the posterior variance is at most $\mu(1-\mu)$. Removing the cheap-type
indicator cannot increase the variance of the posterior, so the actual
one-arrival variance is at most $\varepsilon\mu(1-\mu)\le\varepsilon/4$.
The stopped-martingale argument used above yields
\[
 \E\tau_O\ge\frac{4d_I^2}{\lambda\varepsilon}.
\]
These two bounds prove the observed-default claim and complete
Proposition~\ref{prop:duration}. They establish asymptotic orders in the
specified family; they do not assert a pointwise ranking of exit times
under arbitrary type perturbations or endogenous prices.

\section{Additional comparisons}\label{app:comparisons}

\subsection{Costly access without a private signal}

Keep $\mustar$, the cent menu, the outside option, default shares,
$C=0.3$, and the exponential switching-cost marginal, but remove the
signal obtained on comparing. Paying $C$ now buys rival access only.
Beliefs remain fixed. For a default attached to the priced firm, put
$u=R_i(\mustar)-p$ and $v=R_j(\mustar)-r$. If $u\ge0$, expected
revenue is
\[
 \frac p2\{1-H(v-u-0.3)+H(u-v-0.3)\}.
\]
For the relevant rival prices, $v>0$; if $u<0$, the firm makes no
sales. There is no zero-surplus menu equality at $\mustar$.
At the target rival price, the highest alternative static revenues are
$0.450$ for $A$ and $0.535$ for $B$, below the target revenues
$0.455$ and $0.540$.

For every rival price in the cent rectangle, lower prices outside it
have instantaneous revenue at most $0.390$ for $A$ and
$0.523406174$ for $B$. Since demand is weakly increasing in the rival
price, the target rival price is the adverse corner for these bounds.
Assigning the unrestricted continuation ceiling at the first reset,
with $\alpha\le0.0025$, raises the bounds to at most $0.400280908$
and $0.534431028$. Both remain below target revenue. Prices within the
rectangle and above the reservation boundary are handled exactly as in
Appendix~\ref{app:equilibrium}. The completion argument therefore also
supports the cent targets in this fixed-belief economy. This comparison
establishes that unequal prices alone are not evidence of a learning
mechanism. Without the signal, there is no informative route from prior
$1/2$ to $\mustar$; the fixed belief is an initial condition here.

A fully known quality ranking supplies a different check. Retain the
same participation and access technology and the cent target prices.
The low-quality firm then charges more than $0.5$ and makes no sales.
A cut to $0.49$ retains a positive mass of its high-switching-cost default
consumers, giving positive revenue before the next reset. Thus the same
pair cannot be absorbing under known quality either. Revealing quality
also changes the current buyer's information problem, which is why the
independent-quality comparison in Appendix~\ref{app:equilibrium} is used
to isolate the future relevance of purchases. This check does not import
a covered-market benchmark into the participation game.

\subsection{Synchronized reset opportunities}

Replace the independent clocks by one clock of rate $\alpha$ at which
both firms choose prices simultaneously. This preserves each firm's
individual reset frequency. If both use the cent targets in the silent
rectangle, a unilateral deviation leaves a fixed price pair until the
next common reset. The relevant hazard in the conservative deviation
bound is now $\alpha$, rather than $2\alpha$. By
\eqref{eq:hazardmonotonicity}, the existing bound remains sufficient at
$\alpha\le0.001$. At a reset the rival chooses its target; this price
belongs to the rectangle already covered by the verification. The
finite-game completion allows a simultaneous finite-action stage game
at a common reset, and the same compactness argument applies.

A first arrival buying $B$ before that clock rings enters the silent
region with probability $D_B/(1+\alpha)>0$. At the next common reset,
both targets are installed. Synchronized opportunities therefore also
support silent unequal prices. The asynchronous specification affects
the intervening price paths, but is not necessary for existence of this
outcome.

\subsection{A positive switching-cost floor}

In the covered subproblem of Appendix~\ref{app:consumer}, suppose
comparison cost is common, $C=c$, and the switching-cost distribution
has lower endpoint $\kappa\ge0$ and positive mass immediately above it.
If $2c+\kappa<\gamma=(2q-1)\Delta$, equal prices are informative
for every $\mu\in[1-q,q]$. To verify the assertion for
$\mu\in[1/2,q]$, put $g=0$ in the responsive-type condition. Since
$v_-\le0\le v_+$ on this interval, the default-$B$ bound becomes
\[
 \Psi_B=\left[\frac{\gamma-\kappa}{2}
       +\{\Delta-\kappa(2q-1)\}(\mu-\tfrac12)\right]_+.
\]
The imposed inequality makes the first term exceed $c$ and the slope
positive. A positive mass of low-switching-cost default-$B$ consumers
therefore responds to the signal. Symmetry covers $[1-q,1/2]$.
The argument does not require a zero lower endpoint for switching costs.
It is a covered-market support result in a different cost region from
the central $c=0.3$ example.

\subsection{An example with mixed reset prices}

Finally, positive variation in reset prices need not be caused by
learning. Consider the covered economy at fixed belief $\mu=0.9$,
with values $(1.5,0.5)$, $q=0.75$, equal default shares, $c=0.31$,
and exponential switching costs with mean $\sigma=0.05/\log2$.
Let the price menu be $\{0,0.05,0.10,0.15,0.20\}$. All purchases are
covered because prices are below the lowest product value. Here
$\pi_+=0.7$, $\pi_-=0.3$, $v_+=13/14$, and $v_-=1/2$.
The largest possible signal-responsive comparison gain is
$\pi_+(v_+-v_-)=0.3<c$, so comparison never makes the final choice
depend on the signal. Some consumers nevertheless pay for access.

Default-$A$ consumers always stay. Default-$B$ consumers switch to $A$
when $K<0.49-p_A+p_B$, a positive cutoff throughout the menu.
The expected revenues are consequently
\[
 u_A=p_A\left[1-\tfrac12e^{-(0.49-p_A+p_B)/\sigma}\right],\qquad
 u_B=\tfrac12e^{-(0.49-p_A)/\sigma}p_Be^{-p_B/\sigma}.
\]
For $B$, the price-dependent factors at the five prices are
$0,0.025,0.025,0.01875,0.0125$. Thus $0.05$ and $0.10$ are tied
maximizers for every rival price. For $A$, revenue is strictly increasing
on the entire menu interval: its derivative is
$1-\tfrac12e^{-(0.49-p_A+p_B)/\sigma}(1+p_A/\sigma)>0$, with the
subtracted term below $0.033856$.

Let $A$ always reset to $0.20$ and let $B$ independently choose
$0.05$ or $0.10$ with equal probabilities at every own reset. These
are stationary best responses for every positive discount and reset
rate: prices reveal no information, and each firm's prescribed future
choices are independent of the rival's current choice. In the stationary
law, $B$'s reset-price variance is $0.000625$, and the expected squared
price gap is $0.01625$. The example rests on exact finite-menu profit
indifference. It demonstrates a possible mixed-reset outcome, without
a genericity claim or an attribution of that variation to learning.

\clearpage
\section{Computational supplement}\label{app:computational}

\subsection{Purpose and relation to the proofs}\label{supp:purpose}

The calculations establish the numerical inequalities used in Propositions 1
and 2 and check the consumer-choice configuration used in Proposition 3.
They are deterministic. There are no Monte Carlo draws or estimated parameters.
The consumer kernel, continuation ceiling, equilibrium completion, and exit-time
arguments are given in Appendices A--D of the paper. The files accompanying this
supplement retain every lower-price bound used in the equilibrium construction.

For the equilibrium result, let $P_i$ be firm $i$'s target price, let
$R_i(\mu)$ be its expected product value, and let $F_{i,s}$ be fixed-price
revenue averaged with exponential rate $s$. After a deviation, the first reset
has rate $2\alpha$. The analytical argument in Appendix C implies
\[
 V_i^{\mathrm{dev}}\le
 \frac{\rho F_{i,\rho+2\alpha}(\mu)+2\alpha R_i(\mu)}{\rho+2\alpha}.
\]
Interval arithmetic bounds $F_{i,\rho+2\alpha}$ from above along every required
price choice and observation history. Unresolved branches receive valid upper
continuations. A concave majorant tightens these bounds on the finer menus.
The upper value must be strictly below the value $P_i/2$ of staying.
The calculation does not determine a global equilibrium policy outside the
silent region; the analytical completion argument supplies its existence.

The reproduction directory retains three historical folder names.
\path{proofs/stage4} contains the cent-menu verification and its local extension.
\path{proofs/stage5b} contains the three-arrival calculation, cheap-comparison
checks, and adjusted targets on the two finer menus.
\path{proofs/stage5} contains an earlier diagnostic showing why the cent targets
cannot simply be carried over to a finer menu. That diagnostic is not a
counterexample to the adjusted targets in Proposition 2. These paths are kept
stable so that the recorded source hashes can be checked without alteration.

\subsection{Coverage, rounding, and strict inequalities}\label{supp:coverage}

The cent menu requires 2,077 lower-price cases per observation regime.
The half-cent and tenth-cent menus require 8,110 and 199,590 cases,
respectively. For the latter two menus, 7,336 price blocks verified by interval arithmetic
cover all 207,700 cases without missing or duplicating a required tick.
Each block encloses every menu tick between its endpoints. Real-price
gaps between adjacent blocks are irrelevant to this finite-menu claim,
but prevent an inference about a continuous price interval.

The implementation uses 90-digit point arithmetic and 50-digit interval
arithmetic, with saved endpoints rounded outwards to 25 decimal places.
The supplementary code records each rival row, covered own-price range,
upper value, and source hash. The saved-result audit checks complete
coverage and every required strict inequality. A separately written
consumer-choice integration checks selected demand, posterior, and tree
calculations; it is a diagnostic rather than independent verification
of every interval operation. The narrowest fine-menu margin is about
$2.39\times10^{-6}$, so retaining the outward inequalities matters.
The argument establishes the finite-menu result conditional on these
explicit computational bounds, not through a sampled plot of profits.

The homogeneous comparison-cost check covers 100 closed intervals spanning
$[0.295,0.305]$. Each bound covers the entire interval. The separate
heterogeneous-cost check covers all joint laws on its stated, narrower
support, using cutoff envelopes. These exercises keep the initial public
belief fixed and do not establish equilibrium for the altered populations.

The JSON filenames retain the word \texttt{certificate} to preserve source
integrity. Each file contains bounds and the hashes of the programs that
produced them. These are computational proof records, not machine-checked
formal proofs. Hash checking establishes that the supplied program files
match the recorded inputs; it does not independently validate the arithmetic
library or establish the mathematical validity of the algorithm.

\subsection{Reproducing and checking the calculations}\label{supp:reproduce}

The recorded calculation environment is Python 3.12.14 with
\texttt{mpmath} 1.4.1. The point calculations use 90 decimal digits and
the interval calculations use 50. A plain-text requirements file is supplied.
The programs use the standard library and \texttt{mpmath}; no external data
or network connection is needed after installing the dependency.

From the reproduction directory, run:
\begin{verbatim}
python verify_saved.py
python verify_stage8.py
python audit_worst_blocks.py
python generate_figure_data.py
\end{verbatim}
The first command checks the saved coverage and source hashes, independently
integrates the short revenue tree, and checks exact information-rate formulas.
The second also checks the 100 whole cost intervals and the dynamic
independent-quality lower bound used in Appendix C. The third freshly
recomputes the tightest price block for each finer-menu and firm pair.
It uses the original interval engine, so it is a targeted reproduction,
not an independent implementation of outward arithmetic. The last command
regenerates the Figure 2 coordinates from its exact formulas and checks them
against the stored coordinate file. Figure 1 uses the short-tree quantities
checked by the first command.

To regenerate all principal proof records, run the following commands in
the indicated directories. This overwrites the corresponding computed
outputs, so retain a copy of the supplied files for comparison.
\begin{verbatim}
# proofs/stage4
python certify_stage4.py --workers 4
python audit_stage4.py
python check_saved_results.py

# proofs/stage5b
python certify_mechanism.py
python certify_fine_menus_envelope.py --workers 4
\end{verbatim}
The last command writes \path{fine_menu_certificate_v2.json}.
To audit the regenerated fine-menu output, copy that file to
\path{fine_menu_certificate.json} in the same directory, then run
\texttt{python audit\_repair.py}. The supplied files already contain the
audited output under the latter name. The full finer-menu calculation took
roughly half an hour in an earlier run; elapsed time depends on the machine.
The earlier menu-refinement diagnostic can be regenerated by running
\texttt{python hostile\_audit.py} in \path{proofs/stage5}.

The supplied \path{scientific_audit.json} and
\path{targeted_numerical_audit.json} state what was checked.
The manuscript-preparation pass checks the unchanged program hashes and saved
coverage; it does not regenerate every deviation tree. The four tightest
blocks were separately recomputed in the preceding mathematical audit.
The distinction between a saved-bound check and a new tree computation is
preserved in the reports.

\subsection{Information curves and the use of computational tools}\label{supp:figures}

Figure 2 evaluates the exact one-arrival Kullback--Leibler divergences
derived in Appendix D. With signal accuracy $q=3/4$, put $a=2q-1$ and let
$\varepsilon$ denote the cheap-inspector share. The two plotted functions are
\[
 D_{\rm hidden}(\varepsilon)
 =a\varepsilon\log\frac{1+a\varepsilon}{1-a\varepsilon},
\]
and
\[
 D_{\rm observed}(\varepsilon)
 =\frac{a\varepsilon}{2}\left\{
 \log\frac{q}{1-q}
 +\log\frac{1-\varepsilon(1-q)}{1-\varepsilon q}\right\}.
\]
The coordinate generator evaluates these expressions at 91 logarithmically
spaced shares between $10^{-4}$ and $0.05$, using 70-digit arithmetic.
The figure is a plot of these formulas, not a simulated estimate of exit time.
The rate statements in Proposition 3 are established by the analytical
exit-time argument, not by fitting slopes to the plotted points.

\end{document}